\documentclass[useAMS,usenatbib]{mnras}
\usepackage{graphicx}
\usepackage{natbib} \bibpunct{(}{)}{;}{a}{}{,}
\usepackage{color}
\usepackage[varg]{txfonts}
\usepackage{rotating}
 
\def\changed{}
\def\changedA{}
\def\changedB{}

\newcommand\ION[2]{\textup{#1\,\textsc{\lowercase{#2}}}}
  \def\HII{\ION{H}{ii}} \def\HeI{\ION{He}{i}}
 \def\HeII{\ION{He}{ii}}  
   \def\CIV{\ION{C}{iv}}
    \def\NII{\ION{N}{ii}}
  \def\NIV{\ION{N}{iv}}

  \newcommand{\msunpyr}{M_\odot\,\mbox{yr}^{-1}}
 \newcommand{\dint}{\,\mbox{d}} 
\newcommand{\kms}{\ifmmode{\,\mbox{km}\,\mbox{s}^{-1}}\else{km/s}\fi}
\newcommand{\msun}{\ifmmode M_{\odot} \else M$_{\odot}$\fi}
\newcommand{\rsun}{\ifmmode R_{\odot} \else R$_{\odot}$\fi}
\newcommand{\lsun}{\ifmmode L_{\odot} \else L$_{\odot}$\fi}
\newcommand{\zsun}{\ifmmode Z_{\odot} \else $Z_{\odot}$\fi}
\newcommand{\velo}{\ifmmode\varv\else$\varv$\fi}
\newcommand{\vinf}{\ifmmode\velo_\infty\else$\velo_\infty$\fi}
\newcommand{\LHE}{\ifmmode L_\ION{He}{ii} \else $L_\ION{He}{ii}$\fi}
\newcommand{\LCIV}{\ifmmode L_\ION{C}{iv} \else $L_\ION{C}{iv}$\fi}
\newcommand{\LWN}{\ifmmode L_{\rm WN5h} \else $L_{\rm WN5h}$\fi}
\newcommand{\NWN}{\ifmmode N_{\rm WN5h} \else $N_{\rm WN5h}$\fi}
\newcommand{\LWNZ}{\ifmmode L_{{\rm WNh, low} Z} \else $L_{{\rm WNh, low} Z}$\fi}
\newcommand{\NWNZ}{\ifmmode N_{{\rm WNh, low} Z} \else $N_{{\rm WNh, low} Z}$\fi}
\newcommand{\SN}{\ifmmode {\rm SN} \else SN\fi}
\title[Light-travel-time diagnostics in early Supernova flash
spectra]{Light-travel-time diagnostics in early Supernova spectra:
  substantial mass loss of the IIb progenitor of SN\,2013cu through a
  superwind} \author[G.\ Gr\"{a}fener and J.\ S.\ Vink]{G.\
  Gr\"{a}fener$^{1,2}$ and J.\ S.\ Vink$^{1}$\\
  $^{1}$Armagh Observatory, College Hill, Armagh, BT61\,9DG, United Kingdom\\
  $^{2}$Argelander-Institut f\"ur Astronomie, Auf dem H\"ugel 71,
  53121 Bonn, Germany}
\begin{document} 

\date{Received ; Accepted}
 
\maketitle

\begin{abstract}
  The progenitors of type-IIb supernovae (SNe) are believed to have
  lost their H-rich envelopes almost completely in the direct pre-SN
  phase. Recently the first ``flash spectrum'' of a SN\,IIb
  (SN\,2013cu) has been presented, taken early enough to study its
  immediate circumstellar medium (CSM).  Similar to a previous study
  by \citet{gro1:14} we analyse the structure and chemical composition
  of the optically-thick CSM using non-LTE model atmospheres.  For the
  first time we take light-travel time (LTT) effects on the spectrum
  formation into account, which affect the shapes and strengths of the
  observable emission lines, as well as the inferred SN
  luminosity. Based on the new CSM parameters we estimate a lower
  limit of $\sim 0.3\,M_\odot$ for the CSM mass, which is a factor
  10--100 higher than previous estimates.  The spectral fit implies a
  CSM in the form of a homogeneous and spherically symmetric superwind
  whose mass-loss rate exceeds common expectations by up to two orders
  of magnitude.  The derived chemical composition is in agreement with
  a progenitor that has just left, or is just about to leave the
  Red-Supergiant (RSG) stage, confirming the standard picture for the
  origin of SNe\,IIb.  Due to its extreme mass loss the SN progenitor
  will likely appear as extreme RSG, Luminous Blue Variable (LBV), or
  Yellow Hypergiant (YHG).  The direct detection of a superwind, and
  the high inferred CSM mass suggest that stellar wind mass loss may
  play an important role in the formation of SNe\,IIb.
\end{abstract} 

 
\section{Introduction} 
\label{INTRO}

The question how massive stars end their lives is a fundamental
question in Astrophysics, important for the understanding of the
lifecycle of stars and the chemical enrichment of the interstellar
medium throughout cosmic evolution. An important goal is to understand
the connection between observed Supernova (SN) subtypes and their
direct progenitors. Considerable progress in this direction has been
made in the recent years through the direct identification of SN
progenitors in pre-explosion images \citep[cf.][]{sma1:09}.  However,
in many cases the unknown (and potentially peculiar) properties of
direct SN progenitors substantially complicate the interpretation of
the results \citep[e.g.][]{yoo1:12,wal1:12,gra1:12}.

With the recent advent of short-cadence transient surveys such as the
intermediate Palomar Transient Survey (iPTF) a new window for the
study of the connection between SNe and their direct progenitors has
been opened. Early SN 'flash' spectroscopy, within about a day after
explosion, allows the analysis of the direct environment of the SN and
its progenitor before it is overrun by the SN shock front. The first
example of such an observation, taken 15.5\,h after explosion, has
been presented for the type-IIb SN\,2013cu (iPTF\,13ast) by
\citet{gal1:14}, who interpreted the early spectrum as a Wolf-Rayet
(WR) type emission-line spectrum. Later {\changedA \citet{gro1:14}
  performed a detailed spectral analysis using the non-LTE model
  atmosphere code CMFGEN \citep{hil1:98}, and} showed that the
spectrum originates from a dense low-velocity wind or circum-stellar
medium (CSM) that is more reminiscent of a Luminous Blue Variable
(LBV) or Yellow Hypergiant (YHG) progenitor. {\changedA Based on the
  abundances, mass-loss rate, and terminal wind velocity from this
  analysis \citeauthor{gro1:14} also predicted the possible spectral
  appearance of the progenitor.}

The reason why the early spectrum of SN\,2013cu appears similar to a
stellar spectrum, albeit about 5 orders of magnitude brighter, is that
the SN is still enshrouded in an optically thick wind/CSM, i.e., the
SN shock front is not directly observed {\changedA
  \citep{gal1:14,gro1:14}}. In such a case the spectrum formation is
similar to the case of optically-thick stellar winds {\changedA of WR
  stars or LBVs, whose emission-line spectra are dominated by
  recombination processes. Such objects obey scaling relations that
  preserve their optical depth scales and emission-line equivalent
  widths over large luminosity ranges \citep[cf.\ Sect.\,5.1
  in][]{gra2:13}. Despite the much lower densities in the spatially
  extended wind/CSM of SN\,2013cu, its early spectral appearance is
  thus qualitatively similar to WR-type winds with extremely low wind
  velcities, such as the ones discussed by \citet{gra1:15},} enabling
a spectral analysis with methods traditionally used for the analysis
of hot stars.

The presence of such a dense CSM is somewhat surprising, and possibly
highly relevant for the formation scenario of SNe\,IIb. Evidence for
the presence of winds/CSM around SNe\,IIb has been found previously
from nebular H$\alpha$ \citep{chu2:91,pat1:95,hou1:96,mau1:10}, radio
\citep{fran1:98,sod1:06,kot1:06}, and X-ray observations
\citep{sod1:06,nym1:09,che2:10}. The presence of a dense, and in some
cases modulated CSM has been interpreted as the result of
non-conservative binary interaction or modulated winds in binary
systems \citep{ryd1:06}, or variable LBV-type mass loss through a
strong stellar wind in the direct pre-SN stage \citep{kot1:06}. In the
case of SNe\,IIb both forms of mass loss could be relevant for the
removal of most of the H-rich outer layers of the progenitor, and thus
for the weakness of hydrogen that characterises this type of SNe.

The binary scenario for the formation of SNe\,IIb is supported by low
mass estimates for their progenitors, implying low wind mass-loss
rates \citep{woo1:94,ham2:09,maz1:09,sil1:09,ber1:12} and the direct
detection of possible binary partners at SN explosion sites
\citep{pod1:93,ryd1:06,mau1:09,fol1:14,fox1:14}.  On the other hand
the evidence for strong LBV-type mass loss in the direct pre-SN phase
supports single star scenarios as the one presented by
\citet{gro1:13}.

The early observations of SN\,2013cu are providing for the first time
the opportunity to analyse the immediate CSM of an SN\,IIb
(progenitor) directly, to obtain further constraints on the above
scenarios. In the present work we analyse the early emission-line
spectrum of SN\,2013cu with state-of-the-art model atmospheres
(Sect.\,\ref{MODELS}), taking for the first time the influence of
light-travel-time (LTT) effects into account (Sect.\,\ref{LTT}). From
our quantitative spectroscopic analysis (Sect.\,\ref{SNANALYSIS}) we
obtain information about the structure, velocity, and chemical
composition of the progenitors wind/CSM, providing information on the
evolutionary status of the progenitor and the nature of its mass-loss
(Sect.\,\ref{DISCUSSION}).  Our conclusions are summarised in
Sect.\,\ref{CONCLUSIONS}.

\section{Model assumptions}
\label{MODELS}

In this section we describe the underlying model assumptions for the
non-LTE spectral synthesis models used in this work. In
Sect.\,\ref{STDMOD} we summarise the model assumptions for our
standard models, in Sect.\,\ref{PARAMETERS} we discuss the relevant
model parameters, and in Sect.\,\ref{FORMAL} the numerical computation
of the observable spectrum. Finally, in Sect.\,\ref{TDEP}, we discuss
the expected deviations from the standard model due to time dependent
effects which may affect the spectrum formation in the {\changedA
  direct CSM of early SNe.}

The models used here have previously been used extensively for the
modelling and spectral analysis of hot stars and their winds, in
particular for stars with optically thick Wolf-Rayet type stellar
winds.

\subsection{Standard model atmospheres}
\label{STDMOD}

{\changedA The models presented here are model atmospheres simulating
  the excitation conditions and the spectrum formation in the SN
  progenitors dense and optically thick wind. We assume that the SN is
  in a sufficiently early stage so that the SN shock, which is the
  source of the observed SN luminosity, is still located below the
  inner boundary of our model atmospheres at very large optical depth.
  In Sect.\,\ref{INTERACTION} we will discuss in more detail how this
  scenario applies to SN\,2013cu. For our models the SN shock thus
  merely serves as a 'background' energy source which provides the
  observed photon flux. The created photons are diffusing through the
  dense wind material before they reach the lower bundary of our
  atmosphere models. Here we assume that this incident radiation field
  is well thermalised, and that the radiative flux stays constant
  within our model atmospheres, i.e., that the model atmospheres are
  in radiative equilibrium.}

Our models are computed under standard assumptions for expanding
atmospheres.  These are spherical symmetry, stationarity, and
homogeneity. The radiation field is computed in the co-moving frame of
reference (CMF) following the method described by \citet{koe1:02}. The
atomic population probabilities $n_i(r)$ are computed in non-LTE
simultaneously with the radiation field under the assumption of
statistical equilibrium (SE). The models take thousands of energy
levels and millions of line transitions into account, taking advantage
of the ``super-level'' approach \citep[cf.][]{gra1:02}. The
temperature structure is computed under the assumption of radiative
equilibrium (RE) using the Unsoeld-Lucy type iteration scheme of
\citet{ham1:03}.

\subsection{Model parameters}
\label{PARAMETERS}

The models are defined by the radiative SN luminosity, and the
structure and chemical composition of the CSM. The luminosity $L_\SN$
and inner-boundary radius $R_\SN$ of {\changedA our models} are
connected with the core temperature $T_\SN$ via Stefan Bolzmann's law
\begin{equation}
  \label{TSTAR}
  L_\SN = 4\pi\,R_\SN^2\,\sigma_{\rm SB}\,T_\SN^4.
\end{equation}
We note that the inner boundary of our models is located at an
(LTE-continuum) Rosseland optical depth $\tau_{\rm R}=20$, i.e.,
$T_\SN$ may differ substantially from a classical effective
temperature $T_{\rm eff}$. $T_{\rm eff}$ is usually defined with
respect to the radius $R_{2/3}$ where $\tau_{\rm R}=2/3$.  {\changedA
  For the CSM from extended high-density stellar winds} $R_{2/3}$ can
be much larger than $R_\SN$.

The density $\rho(r)$ and velocity $\varv(r)$ {\changedA throughout our
  models} are assumed to originate from the homogeneous and stationary
stellar wind of the SN progenitor, with a mass-loss rate
$\dot{M}$. They are connected through the equation of continuity
\begin{equation}
\label{CONT}
  \dot{M} = 4\pi\,\rho\varv r^2.
\end{equation}
We further adopt a $\beta$-type velocity law of the form
\begin{equation}
\label{BETALAW}
v(r)=v_\infty\;\left(1-\frac{r_0}{r}\right)^\beta
\end{equation}
with the terminal velocity $v_\infty$, and connection radius
$r_0\approx R_\SN$. The latter connects the outer velocity law with an
exponential density structure at the inner boundary. We adopt $\beta =
1/2$ throughout this work. {\changedA The main purpose of the adopted
  velocity structure is to ensure high enough densities and optical
  depths to achieve local thermodynamic equilibrium (LTE) at the inner
  boundary of our models, i.e., in regions that are not directly
  observed. In Sects.\,\ref{COMPARISON} and \ref{WIND} we will further
  discuss how the adopted velocity and density structure affects the
  interpretation of our results.}

Finally, the chemical composition is given by mass fractions $X_{\rm
  H}$, $X_{\rm He}$, $X_{\rm C}$, $X_{\rm N}$, $X_{\rm O}$, $X_{\rm
  Si}$ and $X_{\rm Fe}$ of hydrogen, helium, carbon, nitrogen, oxygen,
silicon and the iron-group elements. The iron-group elements (Sc, Ti,
V, Cr, Mn, Fe, Co, Ni) are combined in one generic model atom assuming
relative solar abundances \citep[cf.][]{gra1:02}. The detailed model
atoms used in this work are identical to the ones from
\citet{gra1:08}. We further assume that all line profiles are
intrinsically broadened with a Doppler velocity $\varv_D=10$\,km/s due
to micro-turbulence.

\subsection{The formal integral}
\label{FORMAL}

To compare the obtained models with observations their emergent
spectra are computed in the observers frame. To this purpose
the radiation field is computed ray-by-ray in a cartesian $(p,z)$
coordinate system, where the $z$-coordinate is directed along each ray
towards the observer, and $p$ denotes the impact parameter
perpendicular to $z$. These coordinates are connected to the radius
via $p^2 + z^2 = r^2$.

Based on the obtained atomic population probabilities $n_i(r)$ the
emission and absorption coefficients $\eta_\nu(p,z)$ and
$\kappa_\nu(p,z)$ are computed for each obervers-frame frequency
$\nu$, and the emergent intensity $I_\nu(p)$ is integrated along each
ray via
\begin{equation}
  \label{FORMALEQ}
  I_\nu(p) = \int_{-\infty}^\infty \eta_\nu e^{-\tau(p,z)} \dint z
\end{equation}
where
\begin{equation}
  \tau(p,z) = \int^{\infty}_z \kappa_\nu(p,z') \dint z'.
\end{equation}

This formal integral includes further improvements over the previous
CMF computation, including more complex atomic models with
fine-structure splitting, and thermal frequency re-distibution of
photons due to free-electron scattering \citep[cf.][]{hil1:84}.

\subsection{Time dependence}
\label{TDEP}

In this section we wish to give a short overview of how time-dependent
effects in transient systems may render the assumptions of the
standard model, as outlined above, invalid.

Time dependence may affect our models in two ways. 1) A time-dependent
radiation field may affect the SE and RE equations for which we assume
stationarity in our models. As a result the obtained populations
$n_i(r)$ and temperatures $T(r)$ may change. 2) Light-travel-time
(LTT) effects may affect the radiative transfer, and thus change the
internal and emergent radiation fields.

1) Time dependent terms in the SE and RE equations become improtant in
cases where the radiation field changes faster (on timescale
$\tau_{\rm rad}$) than individual atomic processes. Typically, the
slowest processes involved in the spectrum formation are recombination
processes, and it is important to test whether these are affected. 2)
LTT effects become important when the dimensions of the
spectrum-formation region are so large that the light-travel time
$\Delta r/c \gtrsim \tau_{\rm rad}$.

The fully time-dependent radiative transfer problem and its
application for the spectrum formation in SNe is outlined by
\citet{hil1:12}. However, in their practical computations also these
authors set the time derivatives of the radiation field to zero, which
is equivalent to the neglect of LTT effects. While this is justified
for SN ejecta in later phases it will be problematic for early phases
where $\Delta r/c \gtrsim \tau_{\rm rad}$.

In this work we will concentrate specifically on the latter case and
discuss the role of LTT effects on the spectrum formation in early SN
CSM, while retaining the assumption of stationarity in the SE
and RE equations.  This means that we compute the detailed atomic
populations $n_i(r)$ and the temperature $T(r)$ in the usual manner,
using stationary atmosphere models. Only in a second step, when the
emergent spectrum is computed in the observers frame through
Eq.\,\ref{FORMALEQ}, we take LTT effects into account. This approach is
justified if $\tau_{\rm rad}$ is long enough to ensure a
quasi-stationary state on a radial shell within the CSM that can
be characterised by a single luminosity $L_\SN$.  In a first-order
approximation, neglecting e.g.\ multiple scattering processes, such a
shell (with a time-dependent radius $R_L$) will move throughout the
CSM with the speed of light $c$, and the $n_i(R_L)$ and $T(R_L)$
can be taken from a stationary model with the corresponding luminosity
$L$.

\section{Light-travel-time effects}
\label{LTT}

In this section we discuss the implementation of LTT effects in our
models (Sect.\,\ref{DELTAT}), and their effects on the emergent fluxes
and line profiles in early SN flash spectra (Sect.\,\ref{HEPROFILE}).
As discussed above, such effects are expected to play a role for cases
where the LTT throughout the CSM becomes comparable or larger
than the timescale of the brightness variation $\tau_{\rm rad}$.

As an example, for their early flash-spectrum of SN\,2013cu
\citet{gal1:14} estimated that the initial SN explosion took place
$t_{\rm exp} = 15.5$\,h before the spectrum was observed, i.e.,
$\tau_{\rm rad} < 15.5$\,h. On this timescale light travels $R_{\rm
  exp} = c \times t_{\rm exp} \sim 1.7\,\times\,10^{15}\,\mbox{cm}
\sim 24000\,R_\odot$. Radiation that is emitted at $r=R_{\rm exp}$
will thus only be visible for the observer if it stems from parts of
the CSM that are directed towards us.  The far side of the
CSM is subject to a time delay $\Delta t > t_{\rm exp}$
corresponding to a time before the SN explosion.

Based on the r-band photometry from \citet{gal1:14} we estimate a
brightness change of $f_m \sim 1.7$\,mag/day for the flash spectrum of
SN\,2013cu at the time of observation.  In our model from
Sect.\,\ref{ANALYSIS} we adopt an inner boundary radius of
2300\,$R_\odot$, and an outer boundary radius of $\sim 3 \times
10^5\,R_\odot$. The light-travel time $t_{\rm LTT}$ for these
distances is 0.06 and 8 days respectively. We thus expect that
spectral features which are formed in the outer portion of the CSM
will be substantially affected by LTT effects.

\subsection{Computation of the differential time delay $\Delta t$ and
  brightness change $\Delta m$}
\label{DELTAT}

\begin{figure}
  \parbox[b]{0.45\textwidth}{\center{\includegraphics[scale=0.28]{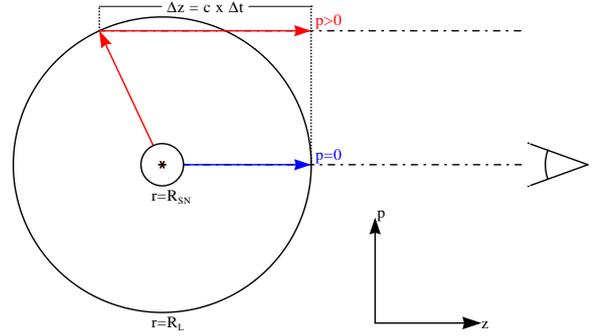}}}
  \caption{\changedB Computation of the time delay $\Delta t$ on a
    radial shell with $r=R_L$. Indicated are two rays in the $(p,z)$
    coordinate system. The origin with $p=z=r=0$ is indicated by the
    asterisk. For the ray with $p>0$ (in red) the time delay
    originates from the indicated distance $\Delta z = c \times \Delta
    t$ (cf.\ Eq.\,\ref{DELTA}). For the central ray with $p=0$ and
    $z=r$ (in blue) $\Delta z = \Delta t =0$.}
  \label{GEOMETRY}
\end{figure}

As outlined in Sect.\,\ref{TDEP} we assume that, at a given time $t$,
the state of the CSM on a radial shell $R_L$ can be described by a
stationary model with luminosity $L_\SN$. Furthermore we assume that
$R_L$ moves in radial direction with the speed of light $c$.
{\changedB In Fig.\,\ref{GEOMETRY} we illustrate the light-travel
  paths in the $(p,z)$ coordinate system for two rays with $p=0$ and
  $p>0$. Only on the radial ray with $p=0$ and $r=z$ (which is
  pointing directly towards the observer) the formal integral
  (Eq.\,\ref{FORMALEQ}) can be computed in the normal way.  Because
  $r=z$, $R_L$ travels simultaneously with the light signal along the
  $z$ axis.  For all other rays with $p > 0$ and $r > z$ the LTT from
  a given radius $r$ towards the observer is longer than for the
  central ray, by the distance $\Delta z$ indicated in
  Fig.\,\ref{GEOMETRY}.}
{\changedB The time difference is
\begin{equation}
  \label{DELTA}
  \Delta t = \frac{\Delta z}{c}= \frac{1}{c}\left[r - z(p,r)\right] 
  = \frac{1}{c}\left[r \pm \sqrt{r^2 - p^2}\right],
\end{equation}
where the minus and plus signs in the last term correspond to the
hemispheres directed towards and away from the observer respectively.}

{\changedB As a consequence the light that reaches the observer from
  an off-center ray with $p>0$ has been emitted earlier by a time
  delay $\Delta t$, compared to the light from the central ray with
  $p=0$.} {\changedA We whish to emphasise that $\Delta t$ is always
  positive. {\changed As long as the SN lightcurve is still rising
    this means that}, compared to the central ray, all other rays have
  time delays $\Delta t$ corresponding to times when the SN was
  fainter.}

We further make the simplifying assumption that the spectrum emitted
at $t-\Delta t$ has the same spectral shape as the one at $t-0$,
except for being fainter by $\Delta m = f_m\,\times\,\Delta t$. Here
$f_m$ denotes the change in magnitude vs.\ time at the time of
observation. Numerically this change is implemented by scaling the
local emission coefficient $\eta_\nu$ in Eq.\,\ref{FORMALEQ} by a
factor $10^{-(2.5\,\times\,\Delta m)}$. {\changedA We note that $f_m$
  characterises the brightness change in a specific wavelength range.
  In reality the spectral shape of the SN will change with time, and
  $f_m$ will be a function of wavelength. $f_m$ should thus be
  estimated for the same wavelength range in which the spectral
  modelling is performed.}

{\changedA As already indicated above, {\changed for early SNe before
    maximum} LTT effects will always lead to a lower brightness than
  the one of the reference model with $L_\SN$.  This will particularly
  affect spectral features that originate from geometrical portions of
  the spectrum-emitting region with large time delays $\Delta t$. If
  also the continuum-emitting region is affected, LTT effects can even
  lead to a decrease of the observed continuum luminosity compared to
  the luminosity $L_\SN$ of the reference model. E.g., for our model
  in Sect.\,\ref{ANALYSIS} we find that the classical photosphere is
  located at $\sim 5\,R_\SN$. The continuum emission thus originates
  from a spherical region beyond $\pm 5\,R_\SN$, i.e., with a spatial
  extension in $z$ that is larger than $10\,R_\SN$. The corresponding
  time delay is of the order of 0.6 days, which implies that the
  continuum emission should be affected by LTT effects.}

\subsection{Theoretical line profiles}
\label{HEPROFILE}

\begin{figure}
  \parbox[b]{0.49\textwidth}{\center{\includegraphics[scale=0.6]{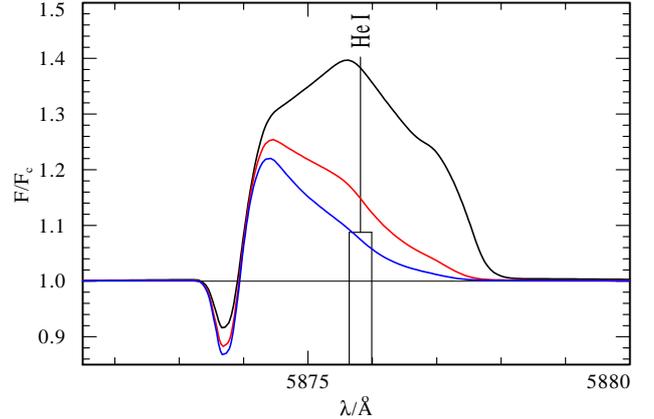}}}
  \caption{The influence of LTT effects on the \HeI\,$\lambda$5876
    line profile. The presented normalised line profiles are computed
    for different adopted brightness changes $f_m = 0.0$ (black), 1.7
    (red), and 3.4\,mag/day (blue).}
  \label{RAWPROFILE}
\end{figure}

In Fig.\,\ref{RAWPROFILE} we present normalised theoretical line
profiles from a CSM model for SN\,2013cu with different adopted
brightness changes $f_m = 0.0$ (black), 1.7 (red), and 3.4\,mag/day
(blue). The model has been computed for a wind velocity of
$\varv_\infty = 103.7$\,km/s. In our analysis in Sect.\,\ref{ANALYSIS}
the presented \HeI\,$\lambda$5876 line is the one which reacts
strongest on LTT effects. The reason is that the line is formed at
large radii far out in the CSM envelope (cf.\,Sect.\,\ref{LINEDEPTH}). For
this reason the unaffected line profile (with $f_m = 0$) nearly
displays a flat-top profile, as expected for optically thin emission
from a hollow shell. However, owing to its small but finite optical
depth, the line also shows a blue-shifted P-Cygni absorption feature
at $- \varv_\infty$, and a moderate blue-red asymmetry due to internal
absorption.

The influence of LTT effects, i.e.\ of an increased $f_m$, on the
emergent spectrum are twofold.  1) {\changedA As discussed in the
  previous section, the light from parts of the CSM that are farther
  away from the observer reaches the observer later, leading to an
  overall fainter spectrum, in particular also in the continuum.}  2)
The red line wings of emission lines in an expanding CSM are
suppressed because they originate from the far side of the CSM which
is moving away from the observer.

 As demonstrated in Fig.\,\ref{RAWPROFILE} this
effect has a strong influence on the normalised line profiles.  For
higher $f_m$ we obtain weaker, narrower, and more blue-shifted line
profiles with a stronger blue-red asymmetry. It is important to note
that these effects are expected to affect lines with different
formation depths differently, and will depend on the adopted value of
$\varv_\infty$. We will discuss the additional diagnostic value of
these LTT effects in Sect.\,\ref{SNPROFILES} for the case of
SN\,2013cu.

\section{The flash spectrum of SN\,2013cu}
\label{SNANALYSIS}

In Sect.\,\ref{ANALYSIS} we present a quantitative spectral analysis
of SN\,2013cu. We note that a similar analysis has been presented
previously by \citet{gro1:14} without taking LTT effects into account.
We will discuss the differences with respect to this work at a later
stage in Sect.\,\ref{COMPARISON}. In Sect.\,\ref{SNPROFILES} we
discuss the additional diagnostic value of LTT effects, followed by a
detailed discussion of the observed H/He (Sect.\,\ref{HHEPROFILES})
and metal line profiles (Sect.\,\ref{METALS}). The final model
parameters and their dependence on the uncertain distance of
SN\,2013cu are discussed in Sect.\,\ref{FINAL}.

\subsection{Quantitative spectral analysis}
\label{ANALYSIS}

\begin{figure*}
  \parbox[b]{0.99\textwidth}{\center{\includegraphics[scale=0.62]{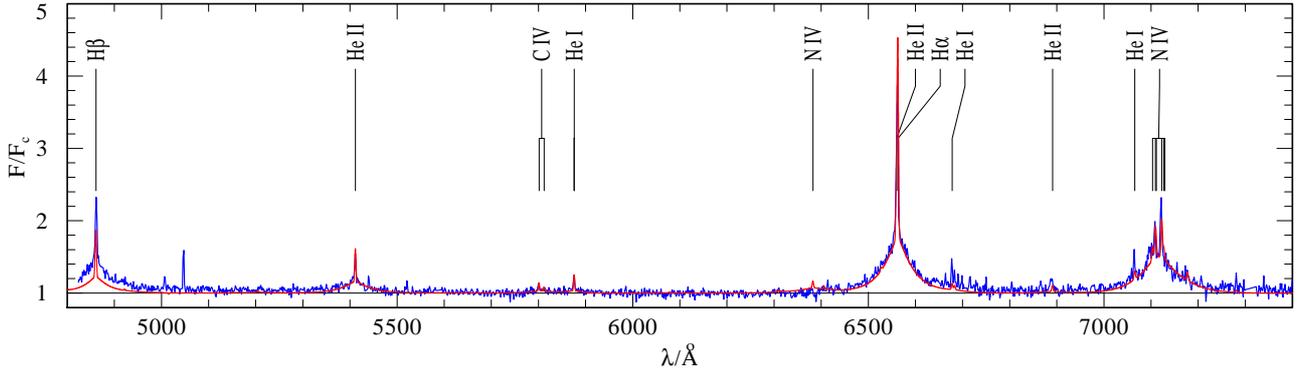}}}
  \caption{Spectral fit for the early flash spectrum of SN\,2013cu,
    15.5\,hrs after explosion.  Normalised spectrum of SN\,2013cu
    (blue) with our best-fit model (red). After correction for
    redshift and extinction, observation and model have been
    normalised in the same way by division through the model
    continuum. }
  \label{FIT}
\end{figure*}

\begin{table}
  \caption{Parameters for the CSM of SN\,2013cu with rough fitting-errors where applicable.}
  \begin{center}
  \begin{tabular}{ll|ll} 
    \hline 
    \multicolumn{2}{l}{\rule{0cm}{2.2ex}{\changedA CSM/wind parameters}} & & {mass fractions} \\
    \hline
    \rule{0cm}{2.2ex}$\log(L_\SN / L_\odot)$    & $10.47\pm0.05$ & X & $0.25\pm0.1$ \\ 
    $\dot{M} / \msunpyr$           & $(4.9\pm0.3)\times10^{-3}$  & Y        & $0.75\pm0.1$ \\
    $R_\SN / R_\odot$               & $4130\pm390$   & Z        & $0.007 {}^{(c)}$ \\
    $T_\SN / {\rm kK}$             & $37.2\pm1.3$   & {\rm C}  & $4.0 \times 10^{-5} {}^{(d)}$  \\
    $\vinf / \kms$                 & $32.8 {}^{(a)}$  & {\rm N}  & $(3.9\pm1) \times 10^{-3}$ \\ 
    $\varv_{\rm D} / \kms$          & $10 {}^{(b)}$    & {\rm O}  & $2.5 \times 10^{-4} {}^{(c)}$ \\
    $\beta$                         & $0.5 {}^{(b)}$   & {\rm Si} & $3.3 \times 10^{-4} {}^{(c)}$ \\
                                    &  & ${\rm Fe} {}^{(e)}$ & $6.3 \times 10^{-4} {}^{(c)}$ \\
    \hline 
    \multicolumn{2}{l}{\rule{0cm}{2.2ex}{\changedA intrinsic model parameters}} \\
    \hline
    \rule{0cm}{2.2ex}$\log(L_\SN / L_\odot)$    & $10.0$  \\
    $\dot{M} / \msunpyr$           & $2.0\times10^{-3}$   \\
    $R_\SN / R_\odot$               & $2405$   \\
    \hline
  \end{tabular}
  \end{center}
  {{\changedA The parameters in the upper part of this table are scaled values that are derived}
    for an adopted distance of 108\,Mpc.
    The systematic uncertainties arising from this assumption and the uncertain value of $\varv_\infty$
    are discussed in Sect.\,\ref{FINAL}. 
    {\changedA The intrinsic model parameters are given in the lower part of the table for the parameters
      that differ from the scaled results.}
    ${}^{(a)}$ This value represents a rough upper limit for $\varv_\infty$
    based on our discussion in Sects.\ \ref{HHEPROFILES} and \ref{FINAL}.
    ${}^{(b)}$ Adopted value.
    ${}^{(c)}$ Indirect estimate based on the expected relative mass fractions of CNO processed material
    relative to the derived N abundance.
    ${}^{(d)}$ Same as (c) but in very good agreement with the observed strength of \CIV\,$\lambda$5801/12.
    ${}^{(e)}$ Combined mass fraction of Fe-group elements with relative abundances from \citet{gra1:02}.}
  \label{TABLE} 
\end{table}

\begin{figure}
  \parbox[b]{0.45\textwidth}{\center{\includegraphics[scale=0.52]{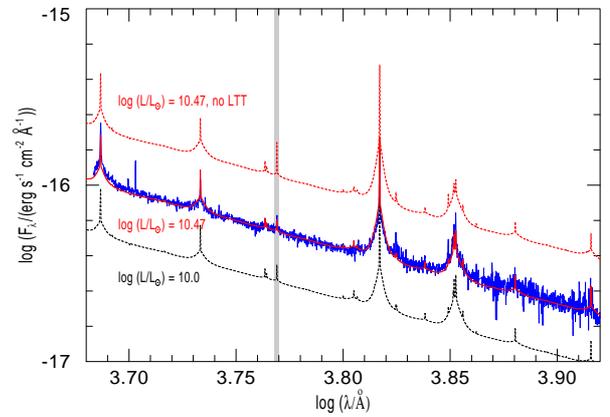}}}
  \caption{\changedA Spectral energy distribution (SED) of SN\,2013cu
    compared to our models. Plotted are the observations in blue, and
    the SED of our best-fit model with LTT effects and
    $L_\SN=10^{10}\,L_\odot$ in black, as well as test models with an
    increased luminosity of $10^{10.47}\,L_\odot$ with and without LTT
    effects in red. Except for the differences in brightness the
    models show different strengths of \HeI\,$\lambda$5876
    (highlighted in grey).}
  \label{SED}
\end{figure}


Our analysis is based on the early flash spectrum of SN\,2013cu as
obtained by \citet{gal1:14} $15.5$\,hrs after explosion with the KECK
DEep Imaging Multi-Object Spectrograph (DEIMOS) grating 600ZD, with a
resolution of $\sim$\,3.5\,\AA\ at 7500\,\AA\ (corresponding to $R\sim
2150$ or a broadening velocity of 140\,km/s).  The data are publicly
available at the Weizmann Interactive Supernova data REPository
\citep[WISeREP,][]{yar1:12}. Following \citeauthor{gal1:14} we scaled
the spectral flux by 1.8\,dex to match their observed host-subtracted
photometry and H$\alpha$ line flux $F_{{\rm H}\alpha} = 3.4 \times
10^{-15} {\rm erg}\,{\rm cm}^{-2}{\rm s}^{-1}$. We further adopt the
same distance of 108\,Mpc, corresponding to a distance modulus $m-M =
35.17$, which is the luminosity distance towards the host galaxy
UGC\,9379 from the NASA Extragalactic Database (NED).

We initially started our analysis with the computation of a model grid
neglecting LTT effects. Using this grid we could match the observed
spectral energy distribution (SED) with a luminosity of
$10^{10}\,L_\odot$ \citep[in very good agreement with the results
of][]{gro1:14}, and an extinction parameter of $E(B-V)=0.1$ using the
standard \citet{car1:88} extinction law with $R_V = 3.1$.

{\changedA 

  The following analysis is based on the same model grid, i.e.\ with a
  fixed luminosity of $10^{10}\,L_\odot$, but includes LTT
  effects. For the brightness change in the optical range we use a
  fixed value of $f_m = 1.7$\,mag/day as estimated from the r-band
  photometry of \citet{gal1:14}. 

  In Fig.\,\ref{FIT} we compare the synthetic line spectrum of our
  best-fit model with observations. To prevent ambiguities due to the
  normalisation procedure the observed spectrum has been normalised by
  dividing the de-reddened obvervation through the model continuum,
  scaled by a constant factor. While this model fits the observed line
  spectrum and the slope of the observed SED very well, the newly
  introduced LTT effects reduce the overall brightness (cf.\
  Sect.\,\ref{DELTAT}), so that the model does not fit the SED in an
  absolute sense anymore. This is illustrated in Fig.\,\ref{SED} where
  we compare the SED of our models, with and without LTT effects, with
  the observed SED of SN\,2013cu.  Our model with $10^{10}\,L_\odot$
  lies now $\sim 0.3$\,dex below the observed SED.  For conventional
  models such a discrepancy can be compensated very precisely with
  common scaling relations (cf.\ Sect.\,\ref{FINAL}).  In the present
  case, however, the inclusion of LTT effects introduces further
  changes in the line profiles and the scaling properties of the
  continuum (cf.\ Sect.\,\ref{FINAL}). In our case these affect
  predominantly the strength of the \HeI\ features and degrade the fit
  quality with respect to the detailed model fits presented below.  To
  illustrate these effects we are plotting a test model with increased
  luminosity, with and without LTT effects, in Fig.\,\ref{SED}.

  The changes introduced by LTT effects complicate the analysis, and
  in particular the determination of the absolute luminosity,
  substantially.  In view of the uncertain distance of the SN host
  galaxy UGC\,9379 and our rather crude approximations concerning the
  inclusion of LTT effects, we decided to restrict our detailed line
  fit to models with a fixed luminosity of $10^{10}\,L_\odot$ in this
  work, to keep the effort for the fitting procedure manageable. Only
  after a satisfactory line fit is obtained we will scale the obtained
  values of $R_\SN$ and $\dot{M}$ to match the observed absolue
  continuum brightness in Sect.\,\ref{FINAL}.

  The line fit in Fig.\,\ref{FIT}} is obtained by adjustingting the
model parameters from Sect.\,\ref{PARAMETERS} to match the diagnostic
emission lines \HeII\,$\lambda$5411, \HeI\,$\lambda$5876,
H$\alpha/\HeII\,\lambda$6562, \CIV\,$\lambda$5801/12, and
\NIV\,$\lambda$7103-29.  The core temperature $T_\SN$, wind density
$\dot{M}/\varv_\infty$, and H/He abundance ratio are determined via
the three available diagnostic features of \HeII, \HeI, and
H$\alpha$/\HeII. When these quantities are fixed the trace element
abundances of C and N can be determined from the \CIV\ and \NIV\
emission lines. The final model parameters are summarised in
Tab.\,\ref{TABLE}. We note that the listed parameters are the ones
obtained after correction of $L_\SN$ for LTT effects {\changedA and do
  not account for systematic differences that may arise from the
  scaling (cf.\ Sect.\,\ref{FINAL})}.

Unfortunately the derived C abundance is very uncertain as the
strength of \CIV\ depends strongly on $T_\SN$. This leaves us in a
situation where the only reliable metallicity indicator left is the
$\NIV$ line. However, even within the large uncertainties, our models
suggest an extremely low C/N ratio $\lesssim 1/50...1/100$ by mass,
implying that the observed material has been processed by the CNO
cycle, and C and O have been almost completely transformed into N (cf.\
our discussion in Sect.\,\ref{EVOLUTION}).  For our final model we
thus decided to use the strength of the \NIV\ line as a proxy for the
total metallicity Z, and to adopt the relative abundances of C, N, O,
Si and the Fe-group elements from an appropriate stellar evolution
model with $Z=Z_\odot$ from \citet{eks1:12}, i.e., to scale their
absolute values with $Z$ as the only free parameter.

So far we determined $T_\SN$, $\dot{M}/\varv_\infty$, and the mass
fractions $X$, $Y$ and $Z$ from the observed emission line strengths
and the shape of the SED. Our synthetic spectrum in Fig.\,\ref{FIT}
reproduces the observed line profiles very well, including the
{\changedA strength of the diagnostic \HeI/\HeII\ lines}, and the
strong electron-scattering wings of H$\alpha$, \HeII\ and \NIV. This
is a big improvement over the work of \citet{gro1:14} who
underestimated the electron-scattering wings and could not reproduce
the observed \HeI\ features. On the downside {\changedA the
  \HeI\,$\lambda$6678, 7065 lines appear} still too weak in our model,
and we predict a weak \NIV\ feature near 6383\AA\ which is not
observed.

In the following we will further constrain the luminosity $L_\SN$, and
the terminal wind velocity $\varv_\infty$. As these values depend
critically on LTT effects we continue here with a detailed dicussion
of LTT effects, specifically for the case of SN\,2013cu. To this
purpose we will have a close look on the narrow cores of the observed
emission lines, and their detailed profiles. Only after $L_\SN$ and
$\varv_\infty$ are fixed we can obtain the mass-loss properties of the
SN progenitor.

\subsection{The influence of LTT effects}
\label{SNPROFILES}

\begin{figure}
  \parbox[b]{0.49\textwidth}{{\includegraphics[scale=0.3]{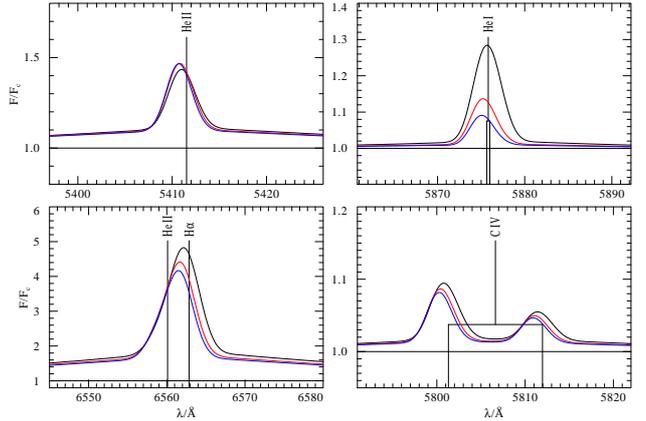}}}
  \caption{Theoretical line profiles for SN\,2013cu assuming different
    brightness changes $f_m = 0.0$ (black), 1.7 (red), and
    3.4\,mag/day (blue). A constant terminal wind velocity
    $\varv_\infty = 103.7$\,km/s is adopted for all models. The
    spectra are convolved with a Gaussian profile corresponding to an
    instrumental resolution of $R=2143$, and a boxcar function
    corresponding to a wavelength sampling of $1.5\,\AA$.}
  \label{PROFILES}
\end{figure}

\begin{figure}
  \parbox[b]{0.49\textwidth}{{\includegraphics[scale=0.3]{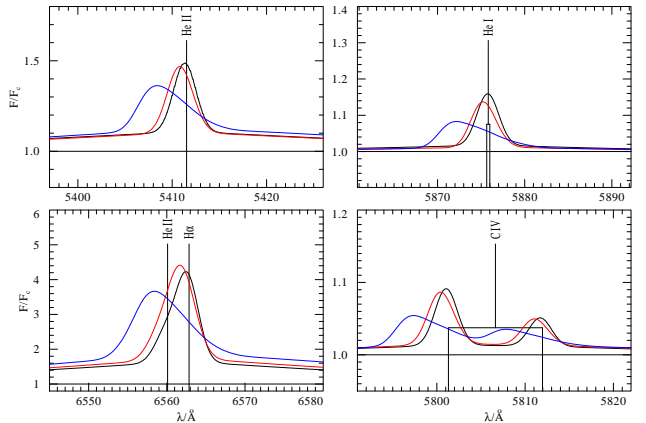}}}
  \caption{Same as Fig.\,\ref{PROFILES} but for a model with fixed
    $f_m = 1.7$\,mag/day, and different terminal wind velocities
    $\varv_\infty = 32.8$ (black), 103.7 (red), and 328.0\,km/s
    (blue).}
  \label{VPROFILES}
\end{figure}

In this section we discuss the specific influence of LTT effects on
the flash spectrum of SN\,2013cu, also to assess their additional
diagnostic value on SN flash spectra in general.  In analogy to
Sect.\,\ref{HEPROFILE} we adopt different brightness changes $f_m =
0.0, 1.7$ and 3.4\,mag/day and investigate their influence on the
spectrum formation for our best-fit model from Sect.\,\ref{ANALYSIS},
which was obtained for $f_m = 1.7$\,mag/day. In particular, we also
take the limited instrumental resolution and relatively coarse
spectral sampling of the observations into account.  To this purpose
we convolve our theoretical spectrum with a Gaussian profile
corresponding to an instrumental resolution of $R=2143$ and a boxcar
function corresponding to a wavelength sampling of $1.5\,\AA$.

As already discussed in Sect.\,\ref{HEPROFILE} we find that, compared
to a cassical model with $f_m=0.0$, our models including LTT effects
are 1) fainter, and 2) show line profiles whose shapes are affected
differently for each line.  For the example of SN\,2013cu we find
substantial brightness reductions by factors of 0.66 for $f_m=1.7$,
and 0.53 for $f_m=3.4$ in the optical range. In our spectral analysis
these effects will increase the derived luminosity roughly by a factor
2, which will also affect the derived size and density distribution of
the CSM envelope.

The effects on the normalised emission-line profiles are shown in
Fig.\,\ref{PROFILES} for a model with a terminal wind velocity
$\varv_\infty = 103.7$\,km/s.  Notably, in Fig.\,\ref{PROFILES},
\HeII\,$\lambda$5411 is hardly affected by the changes of $f_m$
because this high-excitation line is formed near the center of the CSM
envelope. On the other hand, lines that are formed in the outer part
of the CSM, like \HeI\,$\lambda$5876, are strongly affected.  For
these cases an increase of $f_m$ leads to substantially weaker,
narrower and blue-shifted line profiles. For the
H$\alpha/\HeII\,\lambda$6562 complex it appears that the \HeII\
component stays constant while H$\alpha$ is strongly affected. The
\CIV\,$\lambda$5801/12 doublet represents an intermediate case which
is only moderately affected.

The above effects will alter the results of our spectral analysis in
various ways.  The change of the \HeI/\HeII\ ratio will affect the
derived core temperature $T_\SN$. However, as this ratio depends very
sensitively on $T_\SN$, the resulting changes are moderate.  Moreover,
the altered H$\alpha$/\HeII\ ratio will affect the derived He
abundance, while e.g.\ the derived C abundance will remain largely
unaffected (the same holds for the N abundance based on
\NIV\,$\lambda$7103-29).

A very important effect of potentially high diagnostic value is
introduced by the specific wavelength shifts and changes of the line
widths of \HeI\,$\lambda$5876 and H$\alpha$. While individual lines
react differently on changes in $f_m$, the strength of the effect also
depends on $\varv_\infty$. This is demonstrated in
Fig.\,\ref{VPROFILES} where we compare models with different values of
$\varv_\infty$. For this comparison the ratio $\dot{M}/\varv_\infty$
has been kept fixed to ensure the same density distribution for all
models. Otherwise all parameters are kept fixed as listed in
Tab.\,\ref{TABLE}. As to be expected it is clearly possible to
distinguish between wind velocities of the order of 300 and 100\,km/s
as they are larger than the instrumental resolution.  However, from
Fig.\,\ref{VPROFILES} it appears that it may also be possible to
identify line shifts below the instrumental resolution and to
distinguish between terminal wind velocities of the order of 100 and
30\,km/s. {\changedA For observations with high enough S/N the line
  shifts introduced by LTT effects may thus present a new means to
  further constrain $\varv_\infty$, and thus the CSM properties from
  early SN observations like SN\,2013cu.}

\subsection{Hydrogen and Helium line profiles}
\label{HHEPROFILES}

\begin{figure}
  \parbox[b]{0.23\textwidth}{{\includegraphics[scale=0.3]{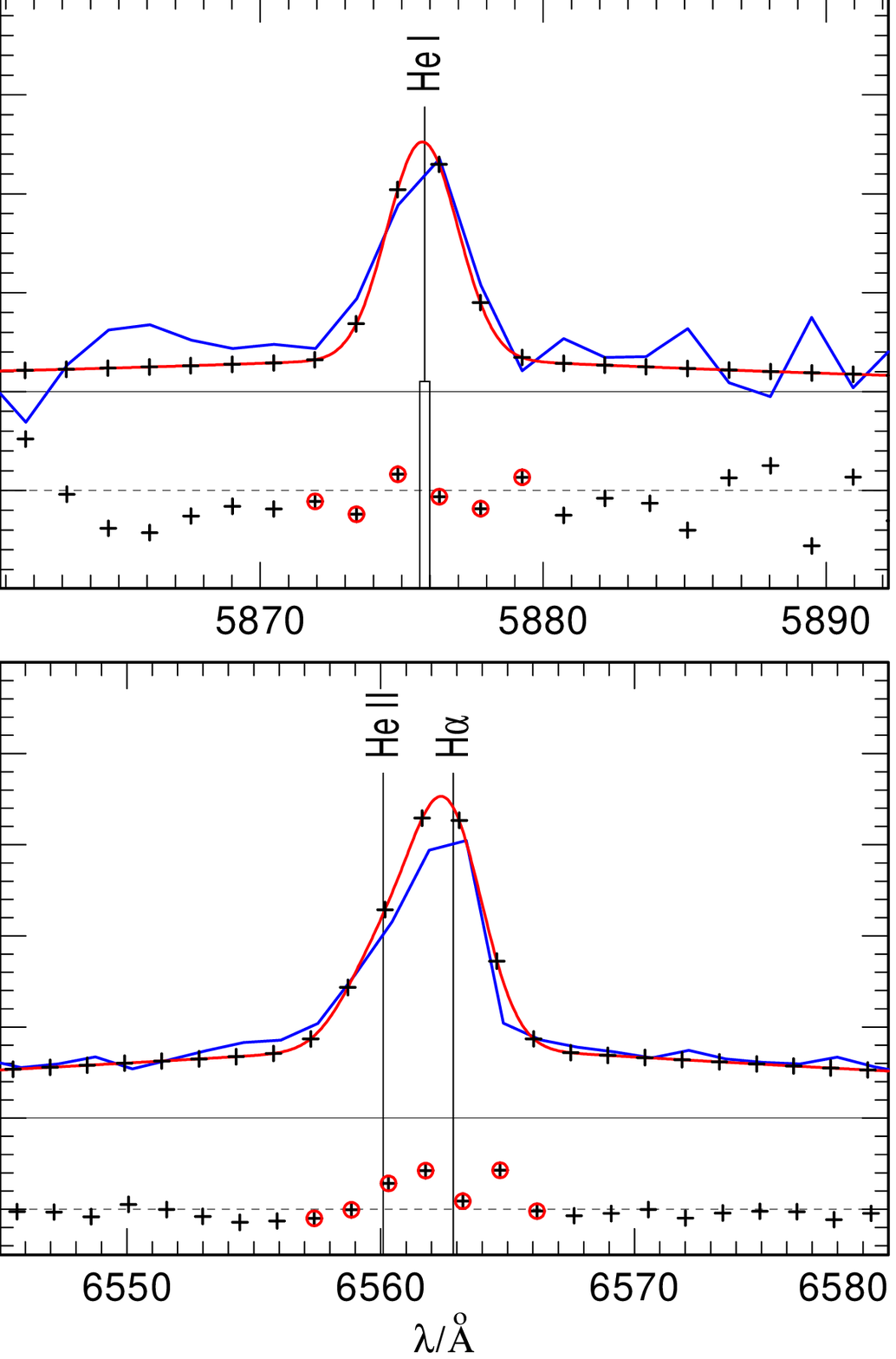}}}
  \parbox[b]{0.23\textwidth}{{\includegraphics[scale=0.3]{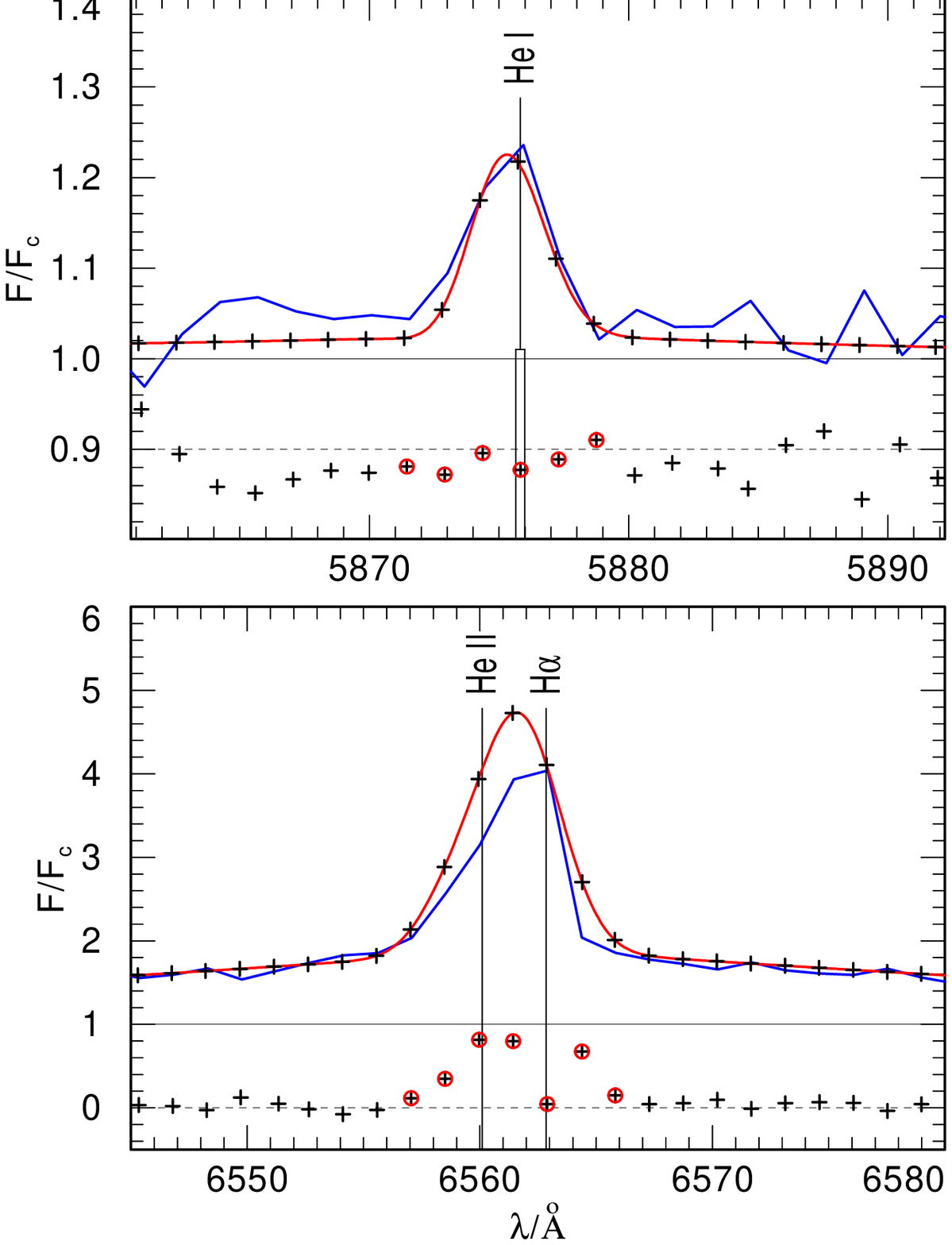}}}
  \caption{Comparison of observed H and He line profiles with models
    in the observers frame. The observed spectra (blue) are corrected
    for redshift to match the emission-line wavelengths as predicted
    by our models (red).  {\changedA Left panel: model with
      $\varv_\infty = 32.8$\,km/s and redshift 7645\,km/s. Right
      panel: model with $\varv_\infty = 103.7$\,km/s and redshift
      7665\,km/s. Residuals are plotted a the bottom of each panel to
      indicate the fit quality (in some cases they are shifted along
      the y-axis). The residuals that are used for the $\chi^2$ plot
      in Fig.\,\ref{HHE_FIT} are encircled in red.}}
  \label{HHE}
\end{figure}

\begin{figure}
  \parbox[b]{0.45\textwidth}{\centering{\includegraphics[scale=0.5]{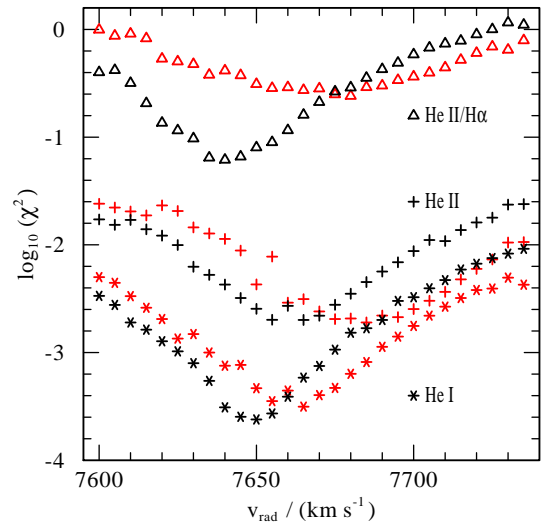}}}
  \caption{\changedA $\chi^2$-like merit function for the three
    diagnostic lines in Fig.\,\ref{HHE}, defined as the average of the
    square of the residuals over the line core (indicated by red
    circles in Fig.\,\ref{HHE}), vs.\ the adopted redshift $\varv_{\rm
      rad}$. Black symbols indicate $\chi^2$ for our model with
    $\varv_\infty = 32.8$\,km/s, red symbols for $\varv_\infty =
    103.7$\,km/s.}
  \label{HHE_FIT}
\end{figure}

In Fig.\,\ref{HHE} we compare the observed line profiles of the narrow
emission-line cores of our three diagnostic H and He lines with
models. In analogy to our comparison in Fig.\,\ref{VPROFILES} we
compare the observations with two models with $\varv_\infty = 32.8$
and 103.7\,km/s and a fixed ratio $\dot{M}/\varv_\infty$. Furthermore,
we adopt the standard brightness change of $f_m = 1.7$\,mag/day.  As
we obtain different intrinsic blueshifts for different values of
$\varv_\infty$ it was necessary to apply slightly different redshift
corrections for the two different models.

{\changedA The derived redshifts are mainly based on the velocity shift
  of \HeI\,$\lambda$5876. In Fig.\,\ref{HHE_FIT} we show $\chi^2$-fits
  for each line in Fig.\,\ref{HHE} as a function of the adopted
  redshift correction $\varv_{\rm rad}$. These help to quantify the
  fit quality and the relative line shifts for the models in
  Fig.\,\ref{HHE}.

  For our model with $\varv_\infty = 32.8$\,km/s (left panels in
  Fig.\,\ref{HHE}, and black symbols in Fig.\,\ref{HHE_FIT}) we obtain
  a good match of the \HeI\,$\lambda$5876 profile simultaneously with
  the H$\alpha$/\HeII\ blend with a redshift of $\varv_{\rm
    rad}=7645$\,km/s within $\pm 5$\,km/s.  Compared to these two
  lines \HeII\,$\lambda$5411 may have a 10--20\,km/s higher redshift.
  In particular, the synthetic H$\alpha$/\HeII\ line profile matches
  the observation well, showing a similar asymmetry as observed.

  For the model with $\varv_\infty = 103.7$\,km/s (right panels in
  Fig.\,\ref{HHE} and red symbols in Fig.\,\ref{HHE_FIT}), the
  obtained fit quality of the H$\alpha$/\HeII\ blend is worse, as
  supported by the higher $\chi^2$ value in Fig.\,\ref{HHE_FIT}.  The
  theoretical profile is symmetrical and does not match the
  observation very well. Furthermore the discrepacy beween the line
  shift of \HeI\,$\lambda$5876 and \HeII\,$\lambda$5411 is larger than
  for our model with low $\varv_\infty$.

  The different symmetries of the theoretical H$\alpha$/\HeII\
  profiles are caused by LTT effects.} While the intrinsic
H$\alpha$/\HeII\ profile is expected to be asymmetrical because of the
different strengths of \HeII\ and H$\alpha$, the asymmetry introduced
by LTT effects (cf.\ Sect.\,\ref{HEPROFILE}) works in the opposite
direction and lets the combined feature appear more symmetrical for
higher wind velocities.  For our model with $\varv_\infty =
103.7$\,km/s these two effects {\changedA appear to} cancel each
other, resulting in an almost symmetrical line profile.  For our model
with $\varv_\infty = 32.8$\,km/s the asymmetry due to LTT effects
becomes smaller, and we obtain a line profile with a similar asymmetry
as observed.

Due to the limited resolution of the present observations the effects
discussed above are subtle, but we think that they are indicative of a
terminal wind speed substantially lower than 100\,km/s. Test
computations with an intermediate value of $\varv_\infty$ still favour
the low value, i.e., $\varv_\infty \lesssim 32.8$\,km/s. Lower values
of $\varv_\infty$ are difficult to assess 1) because of the limited
resolution and S/N of the observations and 2) because of the adopted
line broadening velocity of $\varv_{\rm D} = 10$\,km/s in our models
which introduces qualitative changes in the model structure for lower
$\varv_\infty$.

We conclude that, if we take our predictions for LTT effects
seriously, we favour a value of $\varv_\infty \lesssim
32.8$\,km/s. {\changedA However, we note that this result may still be
  disputable, given the low S/N and coarse binning of the present
  observations.} Based on the width of the observed line profiles
alone we agree with \citet{gro1:14} that values in excess of 100\,km/s
can be firmly excluded because of their substantial effect on the
predicted line profiles (cf.\ Fig.\,\ref{VPROFILES}).

\subsection{Metal lines}
\label{METALS}

\begin{figure}
  \parbox[b]{0.49\textwidth}{{\includegraphics[scale=0.3]{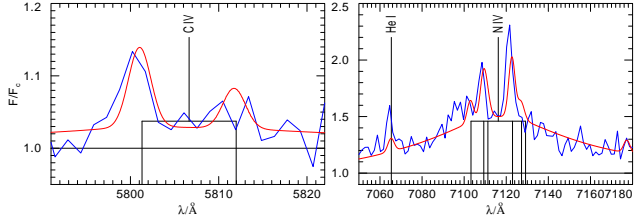}}}
  \caption{Comparison of observed \CIV\ and \NIV\ line profiles with
    our best-fit model ($\varv_\infty = 32.8$\,km/s). The same
    redshift as in Fig.\,\ref{HHE} is applied.}
  \label{NC}
\end{figure}

A detailed comparison of the obseved and predicted \CIV\ and \NIV\
emission lines is shown in Fig.\,\ref{NC}. While the theoretical line
strengths and shapes of these features compare well with the
observations, Fig.\,\ref{NC} reveals clear discrepancies in the
absolute wavelengths. With respect to our synthetic line profiles the
observed \CIV\ doublet is blueshifted by $\sim 30$\,km/s, and the
\NIV\ multiplet by $\sim 50$\,km/s. The reason for this discrepacy is
still enigmatic to us. The most likely reason is that the C and N ions
experience a substantial radiative acceleration due to the intense
radiation field of the SN explosion itself, and decouple from the rest
of the plasma.  This would explain why they are clearly blueshifted
while nearby H and He lines appear unaffected. However, our discussion
in Sect.\,\ref{ARADSN} shows that, despite the extreme radiation field
and the low density in the CSM, the conditions for
ion-decoupling may be difficult to meet.

Particularly for the \NIV\ multiplet it may also be possible that the
wavelengths used in our model atom are incorrect. The wavelengths are
obtained from the atomic database of the National Institute of
Standards and Technology \citep[NIST,][]{kra1:13}, and are based upon
purely theoretical level energies. On the other hand, similar
discrepancies in the \NIV\ wavelengths have never been reported
before. Moreover, the \CIV\ data for which the discrepancies are lower
but still present are also obtained from NIST, but based on observed
wavelengths. We thus think that ion-decoupling is a more plausible
explanation.

\subsection{Final model parameters}
\label{FINAL}

{\changedA As already outlined in Sect.\,\ref{ANALYSIS} our previous
  analysis was based on a model with $L_\SN = 10^{10}L_\odot$ that
  does not match the absolute flux of SN\,2013cu due to the newly
  introduced LTT effects. Our results thus need to be scaled to a
  higher value of $L_\SN$.}

For stars with strong stellar winds such scaling relations are well
established \citep[cf.][]{sch1:89,naj1:97,ham1:98,gra2:13}.  These
relations are based on a scaling of $R_\SN$ and all involved spatial
coordinates with a fixed core temperature $T_\SN$ so that $L_\SN$
follows from Eq.\,\ref{TSTAR}. In the time-dependent case such a
scaling is further complicated by the dependence of LTT effects on
$\Delta r /c \times f_m$. {\changedB As the value of $f_m$ is
  constrained by observations this means that an increase of $L_\SN$,
  and thus $\Delta r$,} leads to even stronger LTT effects so that
$L_\SN$ needs to be increased even further.

{\changedB As demonstrated in Fig.\,\ref{SED} our model with
  $10^{10}\,L_\odot$ under-predicts the optical flux by $\sim
  0.3$\,dex. Due to the non-linearity introduced by LTT effects we
  have to over-compensate this difference substantially, to match the
  absolute optical flux. If we also allow for a variation of the
  extinction coefficient we obtain satisfactory fits in the parameter
  range between $(10^{10.42}\,L_\odot, E(B-V)=0.05)$ and
  $(10^{10.50}\,L_\odot, E(B-V)=0.10)$. In Fig.\,\ref{SED} we show our
  best fit with $(10^{10.47}\,L_\odot, E(B-V)=0.07)$, corresponding to
  an increased radial scale by a factor of $\sim 1.7$.}

{\changedB After correction of $L_\SN$ the obtained mass-loss rate has
  to be scaled with $\dot{M}\propto L_\SN^{3/4} \times \varv_\infty$
  to ensure the preservation of the involved optical depth scales and
  relative emission line strengths \citep[cf.][]{gra2:13}. The values
  of $\dot{M}$ and $R_\SN$ after the scaling are given in the top
  panel of Tab.\,\ref{TABLE}. In the scaled model the now much
  stronger LTT effects lead to a substantial reduction of the strength
  of \HeI\,$\lambda$5876, i.e., to a degradation of the model fit
  (indicated by the grey shaded area in Fig.\,\ref{SED}).} {\changedA
  Unfortunately, LTT effects thus introduce systematic uncertainties
  in our results that will most likely affect the derived H/He
  abundace and $T_\SN$. }

{\changedA As the obtained value of $L_\SN$ also depends on the
  uncertain luminosity} distance towards the SN host galaxy UGC\,9379
we do not attempt to further improve the model fit and use the
obtained parameters in the subsequent discussion as a function
depending on the intrinsic luminosity $L_\SN$.
\begin{equation}
\label{MDOT}
\dot{M} = 4.9 \times 10^{-3} M_\odot \mbox{yr}^{-1} \times \left(\frac{L_\SN}{10^{10.47}L_\odot}\right)^\frac{4}{3}
\times \left( \frac{\varv_\infty}{32.8\,\mbox{km\,s}^{-1}}\right),
\end{equation}
and
\begin{equation}
\label{RSN}
R_\SN = 4130\,R_\odot \times \left(\frac{L_\SN}{10^{10.47}L_\odot}\right)^\frac{1}{2}.
\end{equation}

\section{Discussion}
\label{DISCUSSION}

In the following we discuss the implications of our results on the
nature of the progenitor of SN\,2013cu and its mass loss. {\changedA
  After a discussion of the nature of SN\,2013cu and our model
  assumptions in Sect.\,\ref{INTERACTION}, we compare our results with
  previous works in Sect.\,\ref{COMPARISON}.} In
Sect.\,\ref{LINEDEPTH} we estimate the properties of the ejected
progenitor CSM, and in Sect.\,\ref{ARADSN} we discuss whether our
results are affected by the radiative acceleration of the SN
explosion. Finally, we discuss the properties of the progenitors wind
in Sect.\,\ref{WIND}, and its evolutionary state in
Sect.\,\ref{EVOLUTION}.

{\changedA 
  \subsection{SN\,2013cu as interacting supernova}
\label{INTERACTION}

From the shape of the r-band lightcurve of SN\,2013cu \citet{gal1:14}
estimated an explosion time of $15.5$\,h before the flash spectrum
from Sect.\,\ref{SNANALYSIS} was taken. At this time the SN was still
about 2\,mag below its r-band maximum around day 10, after which the
r-band brightness declined smoothly by $\sim 1$\,mag up to day 30.

The fact that the early spectrum of SN\,2013cu only shows narrow
emission lines with their electron-scattering wings, and no sign of
broad lines that are typically connected with SN ejecta, suggests that
SN\,2013cu is an interacting supernova. In this case the SN luminosity
is created through interaction of the fast SN ejecta with the slow and
optically thick wind/CSM of the progenitor. SNe of this type have been
modeled e.g.\ by \citet{mor1:11} and \citet{des1:15} whose simulations
show that the observed luminosity is created in a thin shock layer
that decelerates while travelling through the CSM.

Assuming an average velocity of $10^4$\,km/s such a shock would travel
$8.6 \times 10^{13}$\,cm or $1241\,R_\odot$ per day. This distance is
much smaller than the inner boundary radius of our models ($R_\SN
\approx 4100\,R_\odot$) and far below the location of the classical
photosphere with $\tau=2/3$ which is located at $R_{2/3}\sim
5.4\,R_\SN$. In agreement with our model assumptions in
Sect.\,\ref{STDMOD} the thin SN shock will thus be located below
$R_\SN$, i.e.\ at very large optical depth, for several days after the
SN explosion. Furthermore the shock will be located below the
photosphere at $R_{2/3}$ for $\sim 2$ weeks.

The spectral evolution of SN\,2013cu is in agreement with this
picture. \citet{gal1:14} provided additional spectra around days 3, 6
and 69. Of these spectra the ones at days 3 and 6 show no sign of
broad emission lines, while the spectrum at day 69 is clearly
dominated by a SN spectrum of type IIb. This implies that around days
3 and 6 the SN shock was still located so far below $R_{2/3}$ that it
was not directly observable. On day 69, on the other hand, the SN
ejecta travelled far beyond $R_{2/3}$, so that the CSM above the SN
shock was optically thin. This late spectrum still shows narrow
H$\alpha$ and [\NII] emission that could be formed in the direct CSM,
or alternatively, in a nearby \HII\ region.

For our upcoming discussion of the observed CSM mass in
Sect.\,\ref{LINEDEPTH} it is interesting whether the SN--CSM
interaction in SN\,2013cu ceased by day 69, as this could set a limit
to the total CSM mass. The presence or absence of strong signs of
interaction \citep[cf.\ also][]{smi1:10} depends on a variety of
parameters, including the density of the CSM, the kinetic energy of
the SN ejecta, and the optical depth of the CSM at the time of
observation. According to \citet{mor1:11} signs of interaction can be
expected for mass-loss rates $\dot{M}>10^{-4}\,\msunpyr$ (assuming
$\varv_\infty=10$\,km/s), which would put SN\,2013cu just above the
limit. {\changedB As \citeauthor{mor1:11} used a high explosion energy
  of $3 \times 10^{51}$\,erg in their computations this limit could be
  higher for the case of SN\,2013cu.} \citet{des1:15} discussed how
the line formation in interacting SNe depends on the location of the
thin SN shock vs.\ the photospheric radius $R_{2/3}$ and showed that,
similar to what is observed for SN\,2013cu, interacting SNe form line
profiles with narrow components in earlier phases when $R_{\rm
  shock}<R_{2/3}$, and with multiple (narrow and broad) components in
late phases when $R_{\rm shock}<R_{2/3}$ and the CSM is optically
thin. We think that it is presently not clear whether the
two-component profiles in the late spectrum of SN\,2013cu are a result
of interaction or not.

{\changedB In any case, based on a shock velocity of $10^4$\,km/s,} an
absence of interaction on day 69 would translate into a maximal
extension of the CSM of $\sim 86000\,R_\odot$ or $21\,R_\SN$. This
vaue is comparable to the size of the \HeI\ line-forming region that
we derive in Sect.\,\ref{LINEDEPTH} and would thus still be compatible
with the CSM mass derived in this section.

Moreover, \citet{mor1:11} showed that the end of the interaction
phase, due to a limited spatial extension of the CSM, would result in
a notable drop in brightness of the SN lightcurve. As the r-band
photometry from \citet{gal1:14} up to day 30 does not show such drop,
the SN--CSM interaction is most likely sustained at least until this
date. Based on the same considerations as above this would indicate a
minimum CSM extension of $2.6\times 10^{15}$\,cm or $\sim
9\,R_\SN$. In any case, an analysis of the lightcurve beyond day 30
could give important clues on the spatial structure of the CSM and the
mass-loss properties of the progenitor of SN\,2013cu.

}

\subsection{Comparison with previous results}
\label{COMPARISON}

The CSM parameters of SN\,2013cu have been estimated previously
by \citet{gal1:14} and \citet{gro1:14}. While \citeauthor{gal1:14}
used analytical estimates of the wind density based on the strength of
H$\alpha$, the analysis of \citeauthor{gro1:14} was based on detailed
spectroscopic modelling very similar to the approach used in the
present work. The main difference is that we take LTT effects into
account, implying a higher luminosity by 0.47\,dex and lower terminal
wind velocity. If we adopt the same luminosity
($\log(L_\SN/L_\odot)=10.0$) and wind velocity
($\varv_\infty=100$\,km/s) as \citeauthor{gro1:14} we obtain a
progenitor mass-loss rate of $4.4\times10^{-3}\,\msunpyr$ from
Eq.\,\ref{MDOT}, which compares well with the estimates of
$5\times10^{-3}\,\msunpyr$ from \citet{gal1:14} and
$3\times10^{-3}\,\msunpyr$ from \citet{gro1:14}, i.e., the results are
largely consistent.

As already noted in Sect.\,\ref{ANALYSIS} the spectral fit obtained in
the present work is superior to the fit presented by \citet{gro1:14}
in two important aspects. 1) we are able to reproduce the strong
electron-scattering wings of H$\alpha$, \HeII\ and \NIV\ very
precisely and simultaneously with the narrow emission line cores of
these lines, while \citeauthor{gro1:14} underpredicted the strength of
the electron-scattering wings. 2) we are able to reproduce {\changedA
  \HeI\,$\lambda$5876 and \HeII\,$\lambda$5411 simultaneously (with
  \HeI\,$\lambda$6678, 7065 weaker than observed)}, while
\citeauthor{gro1:14} could only reproduce \HeII, but no \HeI\
emission.

The electron-scattering wings are an important direct indicator of the
density in the CSM as their strength scales linearly with the electron
density. In comparison, the strength of the narrow emission-line cores
scales with the square of the density because these lines are formed
in emission-line cascades following (two-particle) recombination
processes. In stellar winds the relative strength of these features is
used to determine the degree of small-scale density inhomogeneities
(clumping) within the wind \citep[][]{hil1:91,ham1:98}.  {\changedB
  With respect to a homogeneous, spherically symmetric wind, any
  inhomogeneity increases the mean desity ratio $<\rho^2>/<\rho>$, and
  thus the ratio of the observable emissionline cores vs.\ their
  wings.} The fact that we can reproduce both features simultaneously
{\changedB with a homogeneous wind model thus} indicates that the CSM
likely results from a homogeneous, {\changedB spherically symmetric}
progenitor wind, and not from an inhomogeneous CSM as it may be
expected e.g.\ for mass-loss due to non-conservative binary
interactions.

The abundances derived in the present work are qualitatively similar
to the ones obtained by \citeauthor{gro1:14}, in the sense that we
find a CSM composition substantially enriched in He and N and
depleted in C indicating material that has been processed by the CNO
cycle. In our models we had to reduce the H-abundance to a mass
fraction as low as $X=0.25$ to simultaneously reproduce the observed
strengths of H$\alpha$, \HeI\ and \HeII, while \citeauthor{gro1:14}
derived a higher value of $X=0.46\pm 0.2$, just within the errorbars,
without reproducing \HeI.  Furthermore, our results are in agreement
with a carbon mass fraction of $X_{\rm C}\approx4.0 \times 10^{-5}$
which is four times higher than the value obtained by
\citeauthor{gro1:14}, while our derived nitrogen mass fraction ($X_{\rm
  N}=(3.9\pm1) \times 10^{-3}$) is a factor two lower, indicating a
sub-solar metallicity of $Z \approx Z_\odot/2$.

\begin{figure}
  \parbox[b]{0.49\textwidth}{{\includegraphics[scale=0.57]{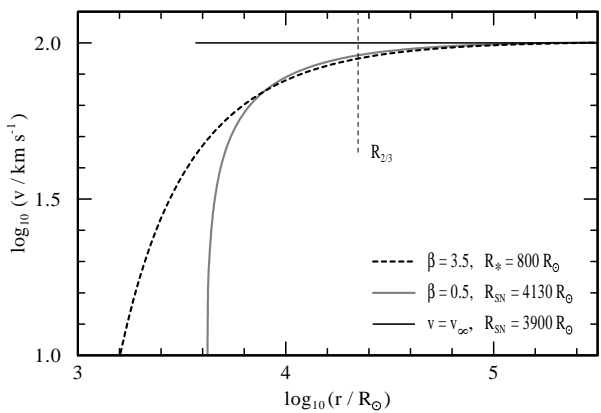}}}
  \caption{\changedA Comparison of our adopted velocity distribution
    with $\beta =0.5$ (grey) with other $\beta$-type velocity laws
    (black dashed line: slowly accelerating RSG wind with $\beta=3.5$;
    thin black line: {\changed constant-velocity CSM model}
    from \citet{gro1:14}). All examples are
    scaled to the same value of $\varv_\infty = 100$\,km/s). $R_{2/3}$
    indicates the photospheric radius with $\tau=2/3$.}
  \label{VELO}
\end{figure}

{\changedB The differences between our work and the work of
  \citeauthor{gro1:14} are most likely caused by the different adopted
  velocity structures, in combination with the temperature sensitivity
  of the involved spectral features. While we use a standard
  $\beta$-type velocity law (Eq.\,\ref{BETALAW}) with $\beta=0.5$,
  \citeauthor{gro1:14} used a {\changed constant velocity with an
    enhaced density at $R_\SN$} to mimic the presence of a thin
  shock. Both velocity structures are shown in Fig.\,\ref{VELO}. For a
  meaningful comparison we scaled the inner radius of the model of
  \citeauthor{gro1:14} using Eq.\,\ref{TSTAR} to match the same value
  of $L_\SN$ as the one derived in this work.
  
  First of all, Fig.\,\ref{VELO} demonstrates that the resulting radii
  $R_\SN$ (and thus the core temperatures $T_\SN$ according to
  Eq.\,\ref{TSTAR}) are rather similar. The main difference lies in
  the velocity and density structure in the innermost part, near and
  below the photospheric radius $R_{2/3}$. At this radius our model
  displays lower velocities, and thus higher densities due to
  Eq.\,\ref{CONT}. Spectroscopically, the higher (electron) densities
  near and slightly below $R_{2/3}$ will affect the strength of the
  electron-scattering wings which are formed in this region (cf.\
  Sect.\,\ref{LINEDEPTH}). The higher densities even even further
  inside will not affect the emergent spectrum as these regions are
  located at large optical depths and are not directly observable.

  The fact that the electron-scattering wings are better reproduced by
  our models can be seen as evidence for higher densities, and thus
  lower velocities near $R_{2/3}$ than adopted by \citet{gro1:14}.
  This indicates a wind acceleration on a large radial scale.
  However, the fact that the region near $R_\SN$ is not directly
  observable also means that the derived value of $R_\SN$ will most
  likely not reflect the actual progenitor radius. In fact a radius of
  $4100\,R_\odot$ is much too large, even for the largest RSGs.
  Instead, the evidence for a slow acceleration near $R_{2/3}$ should
  be interpreted as the result of a very slowly accelerating
  progenitor wind with a much smaller radius.  This is illustrated in
  Fig.\,\ref{VELO}, by a velocity law with $\beta = 3.5$ and
  $R_\star=800\,R_\odot$ that could represent a slowly accelerating
  RSG wind (cf.\ our discussion in Sect.\,\ref{WIND}).  Alternatively,
  a time-dependent, slowly increasing mass-loss rate could be invoked
  to mimick a similar density structure.

}

\subsection{Line-formation depth and CSM mass}
\label{LINEDEPTH}

\begin{figure}
  \parbox[b]{0.49\textwidth}{{\includegraphics[scale=0.38]{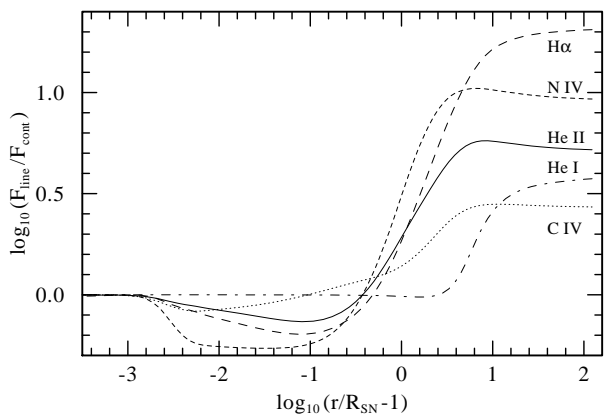}}}
  \caption{Ratio of line and continuum flux for our main diagnostic
    lines \HeII\,$\lambda$5411, \HeI\,$\lambda$5876,
    H$\alpha/\HeII\,\lambda$6562, \CIV\,$\lambda$5801/12 and
    \NIV\,$\lambda$7103-29 as an indicator of the line formation depth
    (see text).}
  \label{DEPTH}
\end{figure}

The fact that our models reproduce the observed \HeI\ features, which
are formed at very large radii, has important consequences for the
derived minimum CSM mass and thus the mass-loss properties of the SN
progenitor.  To assess the amount of material that is directly
observed via line emission we investigate in the following the
formation depths of the diagnostic emission lines in our models.

To this purpose we extract CMF line fluxes from our models.  In the
CMF the profile shapes are qualitatively different from the observers
frame. They consist of a line core near the central wavelength and a
red-shifted line wing that can be either in emission or absorption.
For our present models the extremum (i.e.\ either the maximum or
minimum) of the line wing is located near a redshift of
$\varv_\infty$. In Fig.\,\ref{DEPTH} we plot the flux ratio between
the extremum in the line wing ($F_{\rm line}$) and the continuum just
blueward of the line core ($F_{\rm cont}$) as a function of radius.  Most
lines start with $F_{\rm line}/F_{\rm cont} < 1$ at small radii, i.e., at
small radii these lines are in absorption. Above this region $F_{\rm
  line}/F_{\rm cont}$ increases until a saturation point with $F_{\rm
  line}/F_{\rm cont} > 1$ is reached. We consider this point as the outer
boundary of the formation region of the emission line. Beyond the
saturation radius the line is passive and the line flux marginally
decreases due to absorption.

From Fig.\,\ref{DEPTH} it becomes clear that \NIV\ is the line formed
most deeply in the CSM envelope, followed by H$\alpha$ and \HeII. As the
electron densities in the corresponding layers are high these lines
have the strongest electron-scattering wings. Their formation region
lies clearly below $10\,R_\SN$, i.e.\ near the classical photosphere
with $\tau=2/3$ which is located at $5.4\,R_\SN$ due to the extension
of the CSM. The formation region of \HeI\ lies much further
outward and ends at $\sim 25\,R_\SN$. With $R_\SN$ from Eq.\,\ref{RSN}
a continuous wind would need at least
\begin{equation}
\label{TAUWIND}
25\,R_\SN/\varv_\infty = 69.5\,{\rm yr}
\times \left(\frac{L_\SN}{10^{10.47}L_\odot}\right)^\frac{1}{2} 
\times \left( \frac{\varv_\infty}{32.8\,\mbox{km\,s}^{-1}}\right)^{-1}
\end{equation} 
to reach this radius.
The total mass loss of the SN progenitor during this time is
\begin{equation}
  \dot{M} \times 25\,R_\SN/\varv_\infty = 0.34\,M_\odot
  \times \left(\frac{L_\SN}{10^{10.47}L_\odot}\right)^\frac{11}{6}.
\end{equation} 
We note that this value is a rough lower limit for the total mass-loss
in the direct pre-SN stage as it only refers to the portion of the
CSM which is observed directly. Owing to the larger size of the
\HeI-emitting region compared to H$\alpha$, and the higher luminosity
the value derived here exceeds previous estimates by
\citet{gal1:14,gro1:14} by a factor 10 -- 100.

\subsection{Radiative acceleration during the SN explosion}
\label{ARADSN}

To prepare our discussion of the properties of the SN progenitor we
examine here whether the radiative acceleration due to the SN
explosion itself has an influence on the velocity field in the
observed CSM. In our models the radiative acceleration is
computed explicitly by integrating the product of opacity and flux in
the CMF \citep[cf.][]{gra1:05}. The resulting radiative acceleration
can be written as
\begin{equation}
\label{ARAD}
a_{\rm rad}
= \kappa_F \frac{F_{\rm rad}}{c}
= \kappa_F \frac{L_\SN}{4\pi r^2 c},
\end{equation}
where $F_{\rm rad}$ denotes the radiative flux and $\kappa_F$ the
flux-weighted mean opacity.

For our present models $a_{\rm rad}$ turns out to be remarkably close
(within 0.1\,dex) to the value expected for pure electron scattering,
i.e., $\kappa_F \approx \kappa_{\rm e} = 0.2\times(1 + X_{\rm
  H})\,{\rm cm}^{2} {\rm g}^{-1}$. With $R_\SN$ from Eq.\,\ref{RSN} we
thus obtain
\begin{equation}
  a_{\rm rad}
  \approx \kappa_{\rm e} \frac{L_\SN}{4\pi r^2 c} = \frac{26\,{\rm km\,s^{-1} h^{-1}}
  (1 + X_{\rm H})}{{(r/R_\SN)}^2}.
\end{equation}

As the timescale $\tau_{\rm rad}$ is of the order of several hours we
expect a substantial increase of the velocity near the inner boundary
of our models (where $r \approx R_\SN$).  However, due to the decline
of $a_{\rm rad}$ with $1/r^2$ the velocity increase is only of the
order of few km/s near the photosphere (at $r \approx 5\,R_\SN$) and
negligible further out. In fact, there are signs of slight asymmetries
in the electron-scattering wings in Figs.\,\ref{HHE} and \ref{NC} that
may originate from the additional acceleration in the deep
sub-photospheric layers.

Another important aspect in this context is the possibility of ion
decoupling, i.e., that complex ions experience a stronger radiative
acceleration than H and He due to the larger number of absorbing
spectral lines. If the radiative acceleration of these ions is strong
enough to overcome Coulomb coupling they may separate from the H/He
component of the plasma and show a distinct velocity field. In
Sect.\,\ref{METALS} we discussed that such an effect may indeed be
observed for the \CIV\ and \NIV\ features in SN\,2013cu.

A rough estimate of the importance of ion decoupling in a stellar wind
is given by \citet[][Sect.\,8.1.2]{lam1:99}. They find that the
condition for the efficiency of Coulomb coupling depends on the ratio
of luminosity over wind density via $L\,\varv/\dot{M} \lesssim
5.9\times 10^{16}$ (in $L_\odot\,{\rm km\,s^{-1}}/\msunpyr$).  Using
the paremeters from Sect.\,\ref{PARAMETERS} we obtain
$L\,\varv/\dot{M} \approx 1.5\times 10^{14}$ which suggests that ion
decoupling should not occur.

In any case, we do not expect that the narrow emission components of H
and He are substantially affected, i.e., the velocities derived from
these lines should be representative for the progenitor wind.

\subsection{Properties of the progenitor wind}
\label{WIND}

From our previous analysis and discussion we obtained the following
clues on the mass loss of the progenitor of SN\,2013cu.

\noindent 1) The strong and narrow emission-line cores suggest a high
mass-loss rate of $\sim 5 \times 10^{-3}\msunpyr$ and low terminal
wind velocity, possibly $\lesssim 30$\,km/s.

\noindent 2) The strength of the electron-scattering wings suggests a
homogeneous wind with a smooth density profile, {\changedA that is
  accelerated on a large radial scale}.

\noindent 3) The size of the line-emitting region suggests that the
wind was active over a period of at least 70\,yr before the SN
explosion, and that the mass lost during this period is $\gtrsim
0.3\,M_\odot$.

First of all, the high mass-loss rate and large CSM mass indicate that
the observed pre-SN mass loss may be relevant for the removal of the
progenitors H-envelope and thus for the formation of a SN of type
IIb. {\changedA In principle, a dense CSM could be formed in single
  star scenarios with {\changedB strong} stellar-wind mass loss in the
  direct pre-SN phase, or in binary scenarios with non-conservative
  mass transfer where a fraction of the H-envelope is expelled during
  the interaction process.}

{\changedA In the present case we think that the indications of
  homogeneity strongly support the stellar-wind scenario, because
  binary interaction scenarios would likely involve disk or spiral
  structures, i.e.\ large-scale inhomogeneities that would increase
  the ratio between the observed emission-line cores and their wings
  {\changedB (cf.\ our discussion in Sect.\,\ref{COMPARISON})}.}  We
thus think that the observed CSM has been ejected in the form of an
extremely strong and slow continuous stellar wind.

{\changedA For the mass-loss timescale we estimated a lower limit of
  several decades.} In combination with the extremely high mass loss,
this timescale is reminiscent of S-Doradus type variability {\changedA
  with episodes of strong mass-loss}, as it is observed for Luminous
Blue Variables (LBVs) or Yellow Hypergiants (YHGs)
\citep[cf.][]{smi1:04}. Indications for this type of variable mass
loss have also been found previously in radio lightcurves of
transitional SNe \citep{kot1:06}. However, as the timescale derived in
Eq.\,\ref{TAUWIND} is a lower limit, the observed pre-SN mass loss
could also have occurred on a much longer timescale, {\changedA
  including the possibility of a superwind at the end of the RSG
  phase \citep[cf.][]{yoo1:10}.}

The low terminal wind speed in the CSM of SN\,2013cu provides further
hints on the nature of the progenitor because the mechanical energy of
a stellar wind will be at least of similar order of magnitude as the
energy that is needed to ovecome the stellar gravitational potential,
i.e., $\varv_\infty \gtrsim \varv_{\rm esc}$.  Under this assumption
the radius {\changedA where the stellar wind is accelerated} should be
roughly of the order of
\begin{equation}
  \label{RESC}
  R_{\rm esc} = \frac{2\, G M}{\varv_{\rm esc}^2}
  = 4236\,R_\odot \times \frac{M}{10\,M_\odot}
  \times \left(\frac{\varv_{\rm esc}}{30\,{\rm km/s}}\right)^{-2}.
\end{equation} 

Notably, this value compares very well with our estimate of $R_\SN$ in
Eq.\,\ref{RSN}, {\changedA supporting the large radial scale of the
  wind acceleration in our models (cf.\,Fig.\,\ref{VELO}) in
  accordance with the observed strength of the electron-scattering
  wings (cf.\ our discussion in Sect.\,\ref{COMPARISON}). Slow
  velocity laws such as the one with $\beta=3.5$ in Fig.\,\ref{VELO}
  are found for some RSG winds \citep[e.g.][]{baa1:96,ben1:10}.}

{\changedA However, also} for hotter stars it may be possible to reduce
$R_{\rm esc}$ substantially if the star is located near the Eddington
limit. In this case the mass $M$ in Eq.\,\ref{RESC} may be substituted
by $M_{\rm eff} = M(1-\Gamma_{\rm e})$ where $\Gamma_{\rm e} =
\kappa_{\rm e} L / (4\pi c\,MG)$ is the classical Eddington factor due
to free-electron scattering. In this way a low value of $\varv_\infty$
would point again towards LBV or YHG-like mass loss near the Eddington
limit.  In this case the smooth density structure could result from
S\,Dor-type wind variability that would typically be expected to be of
the order of decades.

What connects LBVs and YHGs with direct SN progenitors is their
proximity to the Eddington limit (cf.\ Sect.\,\ref{EVOLUTION}).
Enhanced mass loss near the Eddington limit has been subject of
several theoretical and empirical studies, for the case of LBV's
\citep{vin1:02}, and more recenly for very massive stars (VMS) near
the top of the main sequence
\citep[][]{gra1:08,gra1:11,vin1:11,vin1:12,bes1:14}. Owing to their proximity
to the Eddington limit, it would be plausible that many direct SN
progenitors exhibit similar mass-loss properties as these objects.

We conclude that the progenitor of SN\,2013cu exhibited exceptionally
strong mass loss over at least several decades in advance of the SN
explosion. The wind properties point towards an unusually strong cool
stellar wind, and/or an enhanced mass-loss rate due to the proximity
to the Eddington limit.  {\changedA {\changedB Based on the inferred
    wind properties} we thus largely agree with \citep{gro1:14} who
  claimed that the SN\,progenitor {\changedB had most likely similar
    spectroscopic/wind properties as LBVs or YHGs}, except that we see
  no reason to exclude the possibility of an RSG progenitor with a
  cool superwind at the end of the RSG stage. In particular, we agree
  that the progenitor of SN\,2013cu was not a WN star, as it was
  originally proposed by \citet{gal1:14}.}

\subsection{Evolutionary status of the SN\,IIb progenitor}
\label{EVOLUTION}

The probably most important result of our analysis is that we confirm
the standard picture for the origin of SNe\,IIb, namely that they are
post-RSG objects that are almost entirely stripped off their H-rich
envelope. This conclusion is based on the following arguments.

Despite the large uncertainties in the C abundance we can say that the
observed value ($X_{\rm C}\approx 4 \times 10^{-5}$) is so low
compared to the solar value \citep[$X_{\rm C}=2.3 \times
10^{-3}$,][]{asp1:09} that carbon must have been destroyed in the CNO
cycles. We can further conclude that the observed material is hardly
mixed with unprocessed material, as this would drastically increase
the C abundance. This is further supported by the high N/C ratio
($X({\rm N})/X({\rm C}) \approx 100$) wich is in very goood agreement
with the CNO equilibrium value \citep[cf.\ right panel of Fig.\,3
in][]{arn1:93}.  Furthermore, the low H abundance ($X=0.25$) indicates
that H-burning in the CNO-cyles is so advanced that also most of the
initial O must have been transformed into N \citep[cf.\ left panel of
Fig.\,3 in][]{arn1:93}. This is further supported by the observed
absolute N abundance ($X_{\rm N} = 3 ... 5 \times 10^{-3}$), which is
higher than the solar carbon abundance. We conclude that nitrogen most
likely provides a very good proxy for the initial CNO abundance, and
thus for the metallicity $Z$ of the SN progenitor.

Based on the arguments above the observed composition likely resembles
pure CNO-cycled material as it is found directly above the H-burning
shell in RSGs \citep[e.g.\ Fig.\,2 in][]{geo1:14}. In agreement with
the standard scenario for SNe\,IIb the observed abundance pattern thus
suggests that most of the H-rich envelope of the SN progenitor has
been removed prior to explosion. An inspection of the evolutionary
tracks of \citet{eks1:12} reveals that only few models show a similar
surface composition at the end of their evolution. In
Tab.\,\ref{EKSTROEM} we compare the stellar parameters of these models
with our results {\changedA \citep[cf.\ also][]{gro2:13}}.

\begin{table}
  \caption{Comparison between rotating and non-rotating single star progenitor models
    at carbon exhaustion from \citet{eks1:12}, and our results for SN\,2013cu.}
  \begin{center}
    \begin{tabular}{lllll} 
        \hline 
        \rule{0cm}{2.2ex} & rotating & non-rot. & non-rot. & SN\,2013cu\\
        \hline
        \rule{0cm}{2.2ex}$M_{\rm ini}/M_\odot$ & 20    & 20   & 25   & - \\
        $M/M_\odot$                    & 7.18  & 8.63 & 8.29 & - \\
        $\log(L / L_\odot)$            & 5.28  & 5.18 & 5.38 & - \\
        $T_{\rm eff} / {\rm K}$        & 16053 & 3715 & 21218 & - \\
        $\dot{M} / 10^{-5}\frac{M_\odot}{{\rm yr}}$    & $1.2 ... 15.3$ & 2.9 & 0.98 & 490 \\
        \hline
        \rule{0cm}{2.2ex}$X$           & 0.24  & 0.48 & 0.16 & 0.24 \\
        $Y$           & 0.75  & 0.51 & 0.83 & 0.75 \\                            
        $Z$           & 0.014 & 0.014 & 0.014 & 0.007 \\
        $X({\rm C})$  & $7.9 \times 10^{-5}$ & $1.0 \times 10^{-4}$ & $8.7 \times 10^{-5}$ & $4 \times 10^{-5}$ \\
        $X({\rm N})$  & $7.9 \times 10^{-3}$ & $6.9 \times 10^{-3}$ & $8.2 \times 10^{-3}$ & $4 \times 10^{-3}$\\             
        $X({\rm O})$  & $5.0 \times 10^{-4}$ & $1.6 \times 10^{-3}$ & $1.1 \times 10^{-4}$ & - \\
        \hline 
        \rule{0cm}{2.2ex} $\Gamma_{\rm e} {}^{(a)}$ & 0.51 & 0.40 & 0.52 & - \\
        \hline
  \end{tabular}
  \end{center}
  {${}^{(a)}$ $\log(\Gamma_{\rm
      e}) = \log(L / L_\odot)-\log(M/M_\odot)+\log(1+X)-4.813$ for a fully ionised plasma.}  
  \label{EKSTROEM} 
\end{table} 

The three models have in common that they just left, or are just about
to leave the RSG stage, i.e., they fit precisely in the standard
picture of SN\,IIb progenitors. Their Eddington factors ($\Gamma_{\rm
  e}$) are in the range 0.4-0.5. For hotter stars semi-empirical
studies of \citet{gra1:11,bes1:14} indicated a strong enhancement of
stellar-wind mass loss in this parameter range (cf.\ also our
discussion in Sect.\,\ref{WIND}). Furthermore, using the mass-loss
recipes of \citet{vin1:02}, \citet{gro1:13} discussed the possible LBV
nature of the rotating progenitor model with $20\,M_\odot$ from
Tab.\,\ref{EKSTROEM} based on its strong and variable mass loss.
Notably, the final mass of this model ($7.2\,M_\odot$) is in agreement
with mass estimates for the progenitor of the broad-lined type IIb
SN\,2003bg by \citet{maz1:09}.

While the qualitative agreement of the observed abundance patterns
with the evolutionary models in Tab.\,\ref{EKSTROEM} is reassuring, it
is important to note that our derived mass-loss rate exceeds
expectations by roughly two orders of magnitude. If this type of
superwind with its extreme mass loss would be persistent enough it may
play an important role in the removal of the H-envelope in the direct
pre-SN stages, and thus for the formation of SNe\,IIb, as well as
stripped-envelope SNe\,Ibc.

\begin{figure}
  \parbox[b]{0.49\textwidth}{{\includegraphics[scale=0.42]{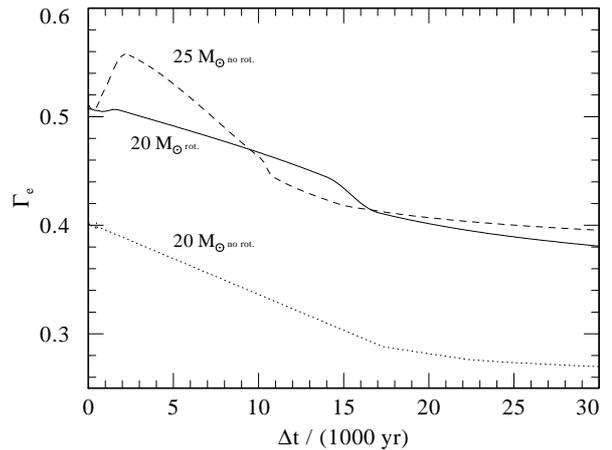}}}
  \caption{Eddington factor for a fully ionised plasma ($\Gamma_{\rm
      e}$) versus time before collapse ($\Delta t$) for the three
    evolutionary models in Tab.\,\ref{EKSTROEM}.}
  \label{GAMMA}
\end{figure}

The proximity to the Eddington limit could be a possible mechanism for
triggering the enhanced mass loss. In Fig.\,\ref{GAMMA} we show
$\Gamma_{\rm e}$ for the evolutionary models in Tab.\,\ref{EKSTROEM}
as a function of time before collapse. Here $\Gamma_{\rm e}$ is
computed for a fully ionised plasma (cf.\ note $^{(a)}$ in
Tab\,\ref{EKSTROEM}). While this assumption holds well for the
atmospheres of stars in excess of $\sim 25000$\,K, for cooler stars
the given values would only be met in regions below the photosphere
where the tempertature is high enough to ionise H and He.  All three
models in Fig.\,\ref{GAMMA} show an increase of $\Gamma_{\rm e}$ from
about 15000\,yr before collapse, which coincides with the time of He
exhaustion in the core. If the observed mass loss is $\Gamma_{\rm
  e}$-driven it could thus persist for several thousand years before
collapse, i.e., much longer than the lower limit of $\sim 70$\,yr from
Eq.\,\ref{TAUWIND}. Consequently the total amount of ejected material
could easily be of the order of several solar masses.

\section{Conclusions and outlook}
\label{CONCLUSIONS}

In this work we analysed the properties of the direct CSM of SN\,2013cu
using a spectrum taken only 15.5\,h after explosion. For the first
time we investigated the impact of light-travel-time (LTT) effects on
the spectrum formation in such an early SN stage.

We found that LTT effects affect the {\changed emergent spectra of
  early SNe (before they reach maximum brightness)} in two
ways. Firstly, the observable spectrum becomes fainter than expected
without LTT effects.  Secondly, the emission-line profiles become
effectively narrower and blue-shifted depending on their formation
depth, i.e., they change differently for different lines.  These
changes also affect the observable line strengths and line ratios. LTT
effects thus provide additional diagnostic means for the quantitative
analysis of early SN spectra.

{\changedA The fact that LTT effects have substantial influence on the
  emergent spectrum and SED, means that they need to be included in
  quantitative spectroscopic studies of objects like SN\,2013cu. In
  view of the involved observational uncertainties and theoretical
  simplifications, this unfortunately also means that potentially
  large systematic uncertainties arise. To get a better handle on the
  detailed time dependence it would be desirable to have high-quality
  high-cadence observations, and at the same time an improved
  numerical treatment of LTT effects.}

{\changedA Our present analysis of SN\,2013cu leads to qualitatively
  similar results as previous works, but with some important
  quantitative differences.}

We confirm the result of \citet{gro1:14} that the SN progenitor was a
star with a strong low-velocity wind, and not a high-velocity
Wolf-Rayet wind as proposed by \citet{gal1:14}. Mainly due to LTT
effects we find a higher mass-loss rate, and indications for a lower
wind velocity $\lesssim 30$\,km/s. From our improved fit of the strong
electron-scattering wings we further conclude that the CSM is likely
homogeneous and spherically symmetric, with a density gradient that
varies slowly on a large radial scale of several thousand $R_\odot$.
These results could indicate a strong cool RSG-like stellar wind, but
could also be interpreted as LBV or YHG-like mass-loss as previously
proposed by \citet{gro1:14}. In any case the observed mass loss is
exceptionally high, up to two orders of magitude higher than what is
adopted in current stellar evolution models.

We further confirm the result of \citet{gro1:14} that the CSM is
deficient in hydrogen and carbon, and enriched in nitrogen, as
expected for CNO-processed material. From our improved fit of the H,
\HeI, \HeII, \CIV, and \NIV\ lines we derive a lower H abundance and
possibly higher C abundance, in very good agreement with expectations
for pure (un-mixed), almost fully CNO-processed material. Such a
composition is predicted for regions directly above the H-burning
shell in RSGs, i.e., it confirms the standard picture that SNe\,IIb
are formed from progenitors whose H-envelopes are almost entirely
removed. {\changedB This may include RSGs at the end of the RSG phase,
  or objects that have just left the RSG stage, with a spectroscopic
  appearance similar to LBVs or YHGs.}

Mainly because of the indications for homogeneity from the precise fit
of the observed electron-scattering wings we believe that the observed
CSM has most likely been formed by a strong superwind, and not through
mass loss in a scenario with non-conservative binary ineraction. This
could indicate that at least a part of the SNe\,IIb is formed in a
single star scenario with strong mass loss in the direct pre-SN stage.

Based on the size of the \HeI-emitting region we deduce an observed
CSM mass of $\sim 0.3\,M_\odot$ which is a factor 10 -- 100
higher than previous estimates based on H$\alpha$. This value poses a
lower limit to the mass lost by the progenitor in the direct pre-SN
stage. Based on the estimated mass-loss rate this corresponds to a
period of strong mass-loss longer than roughly 70\,yr before
explosion. We argue that the occurrence of the exceptionally strong
mass loss could be connected to the proximity of the SN progenitor to
the Eddington limit, and the true CSM mass could be substantially
higher than the derived lower limit.

In the future, an increasing number of early SN observations will
hopefully provide a broader database to examine the properties of
direct SN progenitors, and thus help to connect our understanding of
stellar evolution with the observed types of SNe.  Here we have shown
that LTT effects may provide additional diagnostic means for the
quantitative interpretation of these upcoming observations.  In the
present case we were mainly limited by the low spectral resolution
{\changedA and S/N} of the observations. Moreover, we think that the
theoretical modelling would benefit from a better time coverage with a
cadence of few hours, to get a better handle on the time dependence of
the spectrum.

\section*{Acknowledgments}

We thank the referee, Jose Groh, for his comments that helped to
greatly improve the manuscript. Furthermore we thank T.\ Moriya,
T.-W.\ Chen, and R.\ Kotak for their helpful tips. This work was
funded by the STFC under grant No.\ ST/J001082/1, and the Department
of Culture, Arts and Leisure in Northern Ireland.  This research has
made use of the NASA/IPAC Extragalactic Database (NED) which is
operated by the Jet Propulsion Laboratory, California Institute of
Technology, under contract with the National Aeronautics and Space
Administration.


\end{document}